\documentclass[a4paper,fleqn]{cas-sc}

\usepackage[numbers]{natbib}

\usepackage{amssymb}
\usepackage{amsmath,amsthm}
\usepackage[noend]{algpseudocode}
\usepackage{algorithm} 
\usepackage{array}
\usepackage[caption=false,font=footnotesize]{subfig}
\usepackage{stfloats}
\usepackage{url}
\usepackage{cases}
\usepackage{stfloats}
\usepackage{upgreek}
\usepackage{balance}
\usepackage{color}
\usepackage[dvipsnames]{xcolor}
\usepackage{verbatim}
\usepackage{enumitem,kantlipsum}
\usepackage{threeparttable}
\usepackage{multirow}
\usepackage{rotating,tabularx}

\usepackage{siunitx}
\usepackage{xcolor}
\usepackage[normalem]{ulem}
\usepackage{mleftright}
\usepackage[]{footmisc}
\mleftright

\DeclareMathOperator{\trace}{tr}
\DeclareMathOperator{\diag}{\mathrm{diag}}

\newcommand{\paren}[1]{\left({#1}\right)}
\newcommand{\bracket}[1]{{\left [{#1}\right ]}}
\newcommand{\braces}[1]{{\left\{ {#1}\right\}}} 
\newcommand{\ith}[1]    {{#1}^{\underline{\text{th}}}}
\newcommand{\rr}{_\mathrm{r}}
\newcommand{\cc}{_\mathrm{c}}

\newcommand{\B}{\mathrm{B}}

\newcommand{\target}{\mathrm{t}}

\newcommand{\PiB}{P_{\mathrm{u},i}}

\newcommand{\PBj}{P_{\mathrm{d},mj}}

\newcommand{\HBB}{\mathbf{H}^{\mathrm{SR}}_{\mathrm{u}}}

\newcommand{\sfrac}[2]{#1/#2}

\theoremstyle{definition}

\begin{document}
\shorttitle{Distributed MRMC - III}


\title [mode = title]{Co-Designing Statistical MIMO Radar and In-band Full-Duplex Multi-User MIMO Communications -- Part III: Multi-Target Tracking}    

\tnotetext[1]{K. V. M. acknowledges support from the National Academies of Sciences, Engineering, and Medicine via the Army Research Laboratory Harry Diamond Distinguished Fellowship. The following co-authors are a member of EURASIP: Kumar Vijay Mishra.}

\author[label1]{Sk Nayemuzzaman}
\author[label2]{Kumar Vijay Mishra}[orcid=0000-0002-5386-609X]
\author[label1]{Jiawei~Liu}
\author[label1]{Mohammad~Saquib}

\affiliation[label1]{organization={The University of Texas at Dallas},
            city={Richardson},
            postcode={TX 75080}, 
            country={USA}}
\affiliation[label2]{organization={United States DEVCOM Army Research Laboratory},
            city={Adelphi},
            postcode={MD 20783}, 
            country={USA}}

\maketitle

\begin{abstract}
As a next-generation wireless technology, the in-band full-duplex (IBFD) transmission enables simultaneous transmission and reception of signals over the same frequency, thereby doubling spectral efficiency. Further, a continuous up-scaling of wireless network carrier frequencies arising from ever-increasing data traffic is driving research on integrated sensing and communications (ISAC) systems. In this context, we study the co-design of common waveforms, precoders, and filters for an IBFD multi-user (MU) multiple-input multiple-output (MIMO) communications with a distributed MIMO radar. This paper, along with companion papers (Part I and II), proposes a comprehensive MRMC framework that addresses all these challenges. In the companion papers, we developed signal processing and joint design algorithms for this distributed system. In this paper, we tackle multi-target detection, localization, and tracking. This co-design problem that includes practical MU-MIMO constraints on power and quality-of-service is highly non-convex. We propose a low-complexity procedure based on Barzilai–Borwein gradient algorithm to obtain the design parameters and mixed-integer linear program for distributed target localization. Numerical experiments demonstrate the feasibility and accuracy of multi-target sensing of the distributed FD ISAC system. Finally, we localize and track multiple targets by adapting the joint probabilistic data association and extended Kalman filter for this system.
\end{abstract}


\section{Introduction}
Conventional communications systems are based on either half-duplex (HD) or out-of-band full-duplex (FD) transmission for low-complexity transceiver designs. In such systems, uplink (UL) and downlink (DL) communications are separated in either time or frequency through, for example, time- or frequency-division duplexing, respectively, leading to reduced spectral efficiency \cite{Guo2020FDSI,zhang2016full}. Recently, in response to a tremendous rise in wireless data traffic, in-band full-duplex (IBFD) operation has been suggested as a promising technology to increase spectral efficiency. The IBFD enables concurrent transmission and reception in a single time/frequency channel to potentially double the attainable
spectral efficiency and throughput and reduce latency \cite{song2017full}. 
Recent breakthroughs in SI cancellation (SIC) techniques at both radio-frequency and baseband stages enable an SI suppression of more than $100$ dB and are critical to the more rapid deployment of IBFD-enabled transceivers \cite{Guo2020FDSI}. 
The MIMO evolution of 3GPP Release 18 \cite{3gpp} aggregates two HD sub-bands into a sub-band FD in its schedule.

One of the promising IBFD applications is the emerging area of integrated sensing and communications (ISAC), wherein radar and communications functions concurrently operate in the same spectral range to address the severe spectral crowding problem \cite{mishra2019toward,dokhanchi2019mmwave,duggal2020doppler}. For example, with FD, the base station (BS) is able to receive DL echo signals and estimate radar target parameters. Furthermore, FD allows for the use of joint optimization methods to make a good trade-off. The joint FD ISAC design has the potential for greater spectral efficiency, hardware sharing, and system integration \cite{Schober16distributedFD,singh2018transceiver,FD_WMMSE,Guo2020FDSI,Hassani2020IBFD,Barneto2021FDcommsensing}. 

More recently, the ISAC systems with MIMO radars and MIMO communications (MRMC) have received much attention because, individually, both systems are designed for efficient spectrum usage due to increased degrees-of-freedom in the spatial domain \cite{MCMIMO_RadComm,biswas2018fdqos,Liu2022precoder}. The MIMO radars are usually classified as \textit{colocated} \cite{mishra2019cognitive,Liu2017asilomar} and \textit{widely distributed/statistical} \cite{sun2019target} depending on the geometry of antenna placement. In a colocated MIMO radar \cite{li2007mimo}, the radar cross-section (RCS) is identical to closely-spaced antennas. But, in a distributed MIMO radar, antennas are sufficiently separated and isotropic so that the same target appears with a different RCS to each transmit-receive antenna pair \cite{Liu2018Gloabalsip}. The MRMC literature has largely focused on co-located MIMO radar \cite{sun2019target,sun2024widely}. In the first companion paper (Part I) \cite{liu2024codesigningpart1}, we proposed spectral co-design of statistical MIMO radar with IBFD MU-MIMO communications. In the second companion paper (Part II) \cite{liu2024codesigningpart2}, we developed an algorithm that jointly designed the radar waveform code, communications precoders, and linear receive filters for this distributed system. In this paper, we consider \textit{distributed beamforming} (DB) and multi-target tracking for the same framework.

In wireless sensor networks, DB is employed to achieve the desired signal-to-noise (SNR) and reduce power assumption by providing a coherent beamforming gain \cite{Madhow2010distributed,Brown11multicellsync,Kong2020phaseoffsets,dagefu2020distributedadver,hanna2021distributed,outagedistributed2022}. It is cooperative communications in
which distributed transmitters adjust the phases of their signals
in a way that the signals are constructively combined at a
client \cite{dagefu2020distributedadver}.  Some studies also use this term for beamforming algorithm that is solved in a distributed manner where each user only relies on local information to perform beamforming \cite{Fodor2019distributedhybrid,Valkama2021beamformer}. The coherent combination of various waveforms is accomplished through appropriate synchronization between transmitters \cite{OpenLoop2017Nanzer}. As a mechanism for cooperative communications, distributed transmit beamforming  enables a group of individual source user equipment (UE) to transmit a common message signal as a virtual antenna array such that the bandpass transmissions aggregate constructively after propagation at an intended destination \cite{outagedistributed2022}. 

Compared to conventional beamforming \cite{10056658,10098778,murshed2024self}, the DB relies on each sensor to derive its carrier signal from a separate local oscillator. The independent local oscillator at each UE has random initial phase and phase noise, which forbid phase alignment of signals from different transmitters (Txs) at the receivers (Rxs) of destination UEs \cite{Kong2020phaseoffsets,Madhow2010distributed}. Distributed co-phasing (DCP) is one of the promising techniques to achieve distributed transmit beamforming by combining the coherent gain with the spatial diversity gain. This technique offers benefits such as fixed power transmission from the UEs, robustness to channel estimation errors, and feasibility for practical implementations \cite{chopra2016multistream,outagedistributed2022}. In essence, DCP employs the multiple transmitting nodes as a distributed antenna array to achieve coherent combining gain as well as diversity gain for wireless sensors network \cite{Liu2022DCP}. 

The DB technique has also been explored for multiple-input multiple-output (MIMO) radars in the form of distributed coherent systems, wherein accurate phase synchronization is required to obtain coherent processing gain \cite{Blum2011Coherent,Hongbin2021Sync}. 
Based on the antenna placement relative to the target cross-section (RCS), MIMO radars may be colocated (target RCS appears identical to all Tx-Rx pairs) \cite{li2007mimo,khan2014ambiguity,godrich2010target,boyer2011performance} or distributed (where the Tx and Rx antennas are widely separated and RCS is different for each Tx-Rx pair) \cite{dianat2013target,he2010target,sun2019target}. Distributed MIMO radars may be non-coherent or coherent, depending on whether the phase information is ignored or included \cite{Arye2010sync}. The former extracts diversity gains across different Tx-Rx pairs to overcome target RCS fading, while the latter requires accurate Tx-Rx phase synchronization to exploit processing gains \cite{Blum2011Coherent,Hongbin2021Sync}. Herein, we study the distributed coherent MIMO radar (MIMO radar) because these systems have improved direction finding accuracy over their non-coherent counterparts by ensuring phase coherence of carrier signals from different distributed radar elements \cite{Blum2011Coherent}. 

In practice, the accuracy of the phase in the transmit time slot determines the achievable beamforming gain \cite{Madhow2010distributed}. Therefore, within the realm of DB, considerable efforts have been dedicated to either the carrier frequency/phase synchronization protocol establishment \cite{Brown11multicellsync,Blum2011Coherent,Nanzer2021DPA} or error analysis when mismatched phases occur \cite{Kong2020phaseoffsets,Hongbin2021Sync,outagedistributed2022}. Master-slave ~\cite{Nanzer2021DPA}, round trip~\cite{Brown11multicellsync} and broadcast consensus algorithms~\cite{Blum2011Coherent} are efficient approaches to achieving phase synchronization in a distributed system. The probability of outage with imperfect channel state information (CSI) has also been studied for a distributed wireless sensor network such as cloud radio access network (C-RAN) \cite{outagedistributed2022}. More recently, distributed \textit{co-phasing}, which combines the coherent combining gain with the spatial diversity gain, has been proposed for C-RANs \cite{chaythanya2011comparative,manesh2015physical}. This technique offers benefits such as fixed power transmission from the UEs, robustness to channel estimation errors, and feasibility for practical implementations \cite{chopra2016multistream}.
\begin{table*}
\centering
\caption{Comparison with the state-of-the-art}
\label{tbl:priorcomp}
\begin{threeparttable}
\resizebox{\textwidth}{!}{%
\begin{tabular}{l||c|c||c|c|c|c|c||l}
\hline \multirow{2}{*}{q.v.} & \multicolumn{3}{c||}{Radar}   & \multicolumn{4}{c||}{Communications}  & \multirow{2}{*}{Design objective}  \\\cline{2-7}
& Model & Targets; Clutter & Tracking & Model & Duplexing & Users & Beamformers &   \\
\hline \hline \cite{MCMIMO_RadComm}  & C-MIMO\tnote{a} & Static, single; Yes & No & P2P MIMO\tnote{b} & HD & SU & Max-SINR & Waveforms\\
\hline \cite{he2019performance} & D-MIMO\tnote{c} & Static, single; Yes & No & D-MIMO & HD & SU  &None& Radar Rx filters 
\\
\hline \cite{Lops2020uplink}  & Monostatic & Static, multiple; Yes & No & M-MIMO\tnote{d} & HD (UL) & SU  &Zero Force & Rx filters, BFs\\
\hline \cite{liu2018mimo}  & C-MIMO & Static, single; No & No & MIMO & HD (DL) & MU  & C. I.\tnote{e} & Transmit BFs\\
\hline \cite{biswas2018fdqos}  & C-MIMO & Moving, single; No & No & MIMO & FD &  MU &NSP& BFs, radar waveform \\
\hline \cite{Valkama2021beamformer}  & C-MIMO & Moving, single; No & No & MIMO & FD & MU & Max-SNR  & BF \\
\hline \cite{Zeng2022WaveFDISAC} & Monostatic & Moving, single; No & No & P2P SISO & IBFD & SU & None &  Waveform \\
\hline \cite{liu2024codesigningpart1,liu2024codesigningpart2} & D-MIMO & Moving, single; Yes & No & MIMO & IBFD & MU &Max-MI & Waveform, precoders, filter \\
\hline This paper & D-MIMO & Moving, multiple; Yes & Yes & C-RAN MIMO & IBFD & MU & Co-phased Max-MI & Waveform, precoders, filter, power \\
\hline
\end{tabular}}
\begin{tablenotes}[para]
\item[a] C-MIMO: Colocated MIMO \item[b] P2P: Point-to-point \item[c] D-MIMO: Distributed MIMO \item[d] M-MIMO: Massive MIMO \\\item[e] Constructive interference
\end{tablenotes}
\end{threeparttable}
\end{table*}


\color{black}
Preliminary results of this work appeared in our conference publication \cite{nayem2023multi}, where only a few antenna geometries were considered, DCP was excluded, and optimization algorithm was not described. In this paper, we focus on ISAC design with distributed MIMO radar and IBFD C-RAN, employ co-phasing, use a unified design metric, propose a low-complexity design algorithm, and include multiple targets. Table~\ref{tbl:priorcomp} summarizes our contributions with respect to the state-of-the-art. Our main contributions are:\\
\textbf{1) IBFD C-RAN:} We consider a full duplex C-RAN (FD-C-RAN) where the RRHs are equipped with the IBFD technique and are able to communicate with DMUs and UMUs simultaneously. A typical C-RAN consists of a pool of baseband units (BBUs), a large number of remote radio heads (RRHs), and a Fronthaul network connecting RRHs to BBUs. The BBU pool is deployed at a centralized site, where software-defined BBUs process the baseband signals and coordinate the wireless resource allocation. The RRHs are in charge of RF amplification, up/down-conversion, filtering, analog-to-digital/digital-to-analog conversion, and interface adaption. \\
\textbf{2) Low-complexity design algorithm}: To this end, we employ an alternating minimization procedure, which includes low-complexity Barzilai-Borwein (BB) algorithm \cite{fletcher2005barzilai} for precoder design subproblem. The BB method is an efficient tool for solving large-scale unconstrained optimization problems. 
When compared to the steepest descent method, it has the same search direction but a different step rule. Our BB-based design achieves similar results as the more complex conventional approaches such as the block coordinate descent (BCD) \cite{BCDconvergence}.\\
\textbf{3) Multiple targets:} The presence of multiple targets in a distributed ISAC scenario poses additional challenges. 
Since the relative distance of each target is different with respect to each Rx, the echoes from multiple targets are delayed by a different amount at each Rx. The result is that, after the detection procedure, each Rx ends up with a different ordering of targets in time. This makes it difficult to associate the detected echoes from all Rx uniquely to each target \cite{Behnia2021DataAssociation,Nayebi13dataassociation,KALKAN2014dataassociation,Niesen2019association}. 
To this end, the distributed radar literature suggests various \textit{data association} algorithms such as multiple hypothesis tracking \cite{blackman2004multiple}, random finite sets \cite{mahler2003multitarget}, and joint probabilistic data association (JPDA) \cite{musicki1994integrated}. we propose JPDA to assign detections from both radar and DL signals to specific targets.

The remainder of the paper is organized as follows. The next section describes our FD-ISAC system and the stand-alone radar and communications receivers. In \ref{sec:fd_proc}, we introduce FD D-ISAC receiver processing for  self-interference, radar-to-communications interference, and vice versa. Section~\ref{sec:solution} presents our proposed multi-target CWSM optimization using the low-complexity BBB procedure. The multi-target detection via data association in D-ISAC is discussed in Section~\ref{sec:estimation} followed by extensive numerical experiments in Section~\ref{sec:numerical}. We conclude in Section \ref{sec:conclusion}.

Throughout this paper, lowercase regular, lowercase boldface, and uppercase boldface letters denote scalars, vectors, and matrices, respectively. We use $I(\mathbf{X};\mathbf{Y})$ and $H\paren{\mathbf{X}|\mathbf{Y}}$ to denote MI and conditional entropy between two random variables $\mathbf{X}$ and $\mathbf{Y}$, respectively. The notations $\mathbf{Y}\bracket{k}$, $\mathbf{y}\bracket{k}$, and $y\bracket{k}$ denote the value of time-variant matrix $\mathbf{Y}$, vector $\mathbf{y}$ and scalar $y$ at discrete-time index $k$, respectively; $\mathbf{1}_{N}$ is a vector of size $N$ with all ones; $\mathbb{C}$ and $\mathbb{R}$ represent sets of complex and real numbers, respectively; a circularly symmetric complex Gaussian (CSCG) vector $\mathbf{q}$ with $N$ elements and power spectral density $\mathcal{N}_0$ is $\mathbf{q}\sim\mathcal{CN}(0,\mathcal{N}_0\mathbf{I}_{N})$; $(\cdot)^{\star}$ is the solution of the optimization problem; $\mathbb{E}\bracket{\cdot}$ is the statistical expectation; $\trace\{\mathbf{R}\}$, $\mathbf{R}^\top$, $\mathbf{R}^\dagger$, $\mathbf{R}^\ast$, $\left| \mathbf{R}\right|$, $\mathbf{R}\succeq\mathbf{0}$, and $\mathbf{R}\paren{m,n}$ are the trace, transpose, Hermitian transpose, element-wise complex conjugate, determinant, positive semi-definiteness and $\ith{\paren{m,n}}$ entry of matrix $\mathbf{R}$, respectively; set $\mathbb{Z}_{+}(L)$ denotes $\left\lbrace1,\dots,L\right\rbrace$;  $\mathbf{x}\succeq\mathbf{y}$ denotes component-wise inequality between vectors $\mathbf{x}$ and $\mathbf{y}$; $x^+$ represents $\max(x,0)$; $x^{\paren{t}}\paren{\cdot}$ is the $\ith{t}$ iterate of an iterative function $x\paren{\cdot}$; $\inf(\cdot)$ is the infimum of its argument; $\odot$ denotes the Hadamard product; and $\oplus$ is the direct sum. All distances are measured in kilometers.

\clearpage
\section{
System Model}
\label{sec:system}
\begin{figure}[t]
\centering
\includegraphics[width=1\columnwidth]{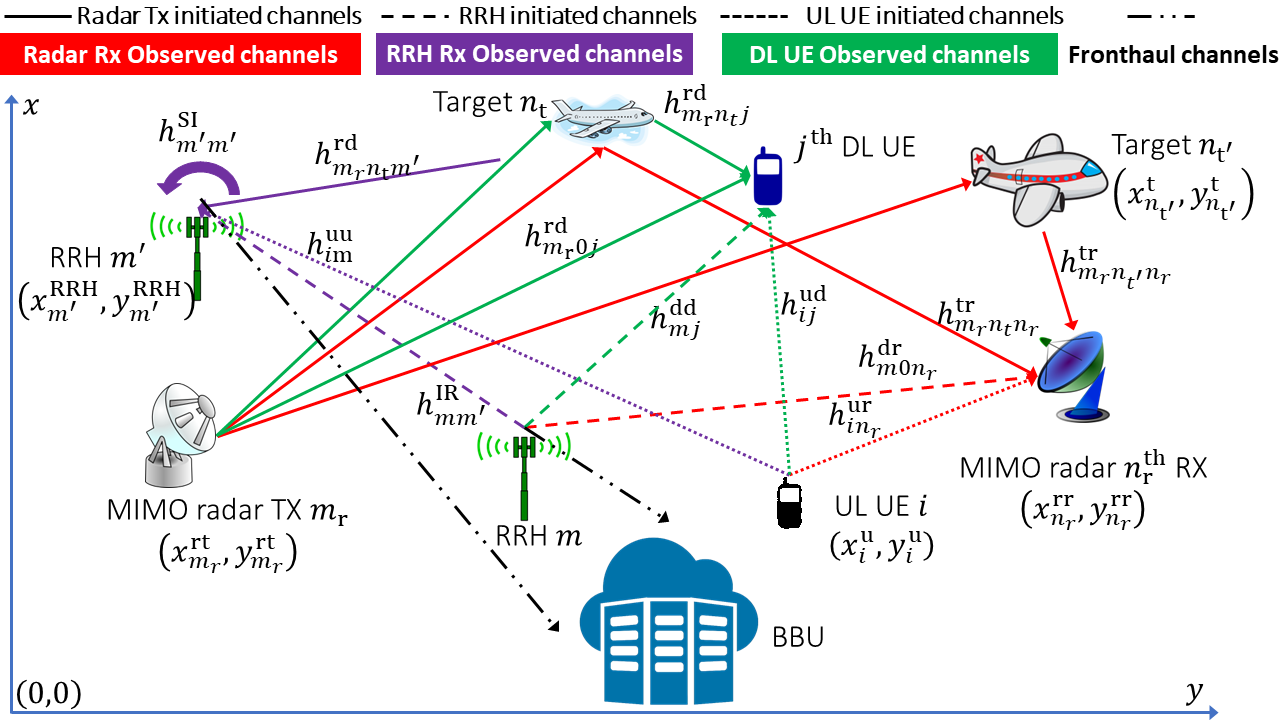}
\caption{Illustration of the considered D-ISAC system, where the $\ith{m\rr}$ ($\ith{n\rr}$) radar Tx (Rx), $\ith{m}$ FD RRH, $\ith{j}$ ($\ith{i}$) DL (UL) UE are located at
$\paren{x^{\mathrm{rt}}_{m_\mathrm{r}}, y^{\mathrm{rt}}_{m_\mathrm{r}}}$ ($\paren{x^{\mathrm{rr}}_{n_\mathrm{r}}, y^{\mathrm{rr}}_{n_\mathrm{r}}}$),  $\paren{x^{\mathrm{RRH}}_{m},y^{\mathrm{RRH}}_{m}}$, $\paren{x^\mathrm{d}_{j},y_{\mathrm{d},j}}$ ($\paren{x_{\mathrm{u},i},y^\mathrm{u}_{i}}$), respectively; the $\ith{n_\target}$ target is located at $\paren{x_{\target,n_\target},y_{\target,n_\target}}$. 
}
\label{fig:setup}
\end{figure}

We consider an FD-ISAC system consisting of a MIMO radar with $M_\mathrm{r}$ ($N_\mathrm{r}$) widely distributed single antenna Txs (Rxs) and an FD C-RAN, which encompasses $N_\B$ FD RRHs jointly serving $\mathit{J}$ ($\mathit{I}$) single antenna HD DL (UL) UEs concurrently. Each FD RRH is equipped with $M\cc$ transceiving antennas and connected to the BBU via a fiber-fronthaul link. The MIMO radar detects and localizes $N_\target$ moving targets within the coverage of the FD-C-RAN during an ISAC operation window when the $M$ FD RRHs coherently broadcast data streams to each DL UE while the $I$ UL UEs multi-access channel to all RRUs. Simultaneously, each radar Tx emits a train of $\mathit{K}$ pulses to detect a moving target in the coverage area of the BS at a uniform pulse repetition interval (PRI) $T_{\mathrm{r}}$; the total duration $KT_{\mathrm{r}}$ is the \textit{coherent processing interval} (CPI). The integration of FD communications and radar sensing allows the radar pulse width, PRI, and CPI to equal the communications symbol duration, frame length, and scheduling window, respectively. As a result, the number of symbol periods in each frame, $L$, equals the number of range bins.  \figurename{~\ref{fig:setup}} illustrates the system model  on a two-dimensional (2-D) $\left(x\textrm{-}y \right)$ Cartesian plane $\left(x,y \right)$. 

\subsection{Transmit signal}
For the FD-C-RAN, we adopt an all-RRH association policy, namely that all corresponding FD RRHs cooperatively transmit DL signals to each DL UE  while each UL UE sends a common single-stream data symbol $M$ FD RRHs \cite{Tellambura2018FDCran}. During the the $\ith{l}$ symbol period of the $\ith{k}$ frame or the $\ith{\bracket{k,l}}$ symbol period, of the ISAC operation window, FD RRH $m\in\braces{1,\cdots,M}$  and UL UE $i\in\braces{1,\cdots,I}$ simultaneously transmit DL and UL signals $x_{\mathrm{d},m}\bracket{k,l}=\sum_{j=1}^J\mathbf{v}_{\mathrm{d},mj}s_{\mathrm{d},j}\bracket{k,l}$ and $x_{\mathrm{u},i}\bracket{k,l}=\sqrt{P_{\mathrm{u},i}}s_{\mathrm{u},i}\bracket{k,l}$, respectively, where $\mathbf{v}^\mathrm{d}_{mj}\in\mathbb{C}^{M\cc}$ and $P_{\mathrm{u},i}$ are the DL beamforming vector and the transmit power employed by the $\ith{m}$ RRH towards the $\ith{j}$ DL UE  and the $\ith{i}$ UL UE, respectively; $s_{\mathrm{d},j}\bracket{k,l}$ ($s_{\mathrm{u},i}\bracket{k,l}$) designates the single the data stream for the $\ith{j}$ DL ($\ith{i}$ UL) UE with $\mathbb{E}\bracket{\lvert s_{\mathrm{d},j}\bracket{k,l}\rvert^2}=1$ ($\mathbb{E}\bracket{\lvert s_{\mathrm{u},i}\bracket{k,l}\rvert^2}=1$).  

Denote the radar pulse duration as $T_\mathrm{p}= T_\mathrm{r}/N_{\mathrm{p}}$ (a.k.a fast time), where $N_{\mathrm{p}}$ is the number of range cells per PRI. We define the radar code vector transmitted during the $\ith{k}$ PRI as $\mathbf{a}\bracket{k}=\bracket{a_{k,1},\cdots,a_{k,\mathit{M}\rr}}^\top\in\mathbb{C}^{M\rr}$ and the MIMO radar code matrix as $\mathbf{A}=\bracket{\mathbf{a}\bracket{1},\cdots, \mathbf{a}\bracket{\mathrm{\mathit{K}}}}\in\mathbb{C}^{M\rr\times K}$. The pulse train transmitted by the $\ith{m\rr}$ radar Tx is written as $s_{m_\mathrm{r}}(t)=\sum_{k=0}^{K-1}a_{k,m\rr}\phi_{m_\mathrm{r}}\paren{t-kT_\mathrm{r}}$ $\mathit{K}$, where $\phi_{m_\mathrm{r}}(t)$ denotes the orthonormal waveform associated with the  $\ith{m_\mathrm{r}}$ radar Tx with support $[0,T_{\mathrm{p}})$. Grouping the transmit signals from $M\rr$ Txs yields $\mathbf{s}(t)=\bracket{s_1(t),\cdots,s_{M_\mathrm{r}}(t)}^\top$.

\subsection{Channel}
\label{sec: system_channel}
During a given CPI, the $\ith{n_\target}$ moving target is located at $\paren{x_{\target,n_{\target}},y_{\target,n_{\target}}}$ with the horizontal and vertical velocity as $\paren{\dot x_{\target,n_\target},\dot y_{\target,n_\target}}$. Then we define the state vector of the $\ith{n_\target}$ target as $\mathbf{x}_{\target,n_\target}=\bracket{x_{\target,n_\target},y_{\target,n_\target},\dot x_{\target,n_\target},\dot y_{\target,n_\target}}^\top$  The propagation delay and Doppler shift associated with the $\ith{\paren{m\rr n_\target n\rr}}$ Tx-target-Rx path are observed as  
\begin{align}
b_{\tau,m\rr n\rr}\paren{\mathbf{x}_{\target,n_\target}}&\triangleq\tau_{m\rr n_\target n\rr}=\frac{r_{\mathrm{r},m\rr n_\target n\rr}}{c}=\frac{r_{m\rr n_\target m}+r_{n_\target n\rr}}{c}
\\
b_{f,m\rr n\rr}\paren{\mathbf{x}_{\target,n_\target}}&\triangleq f_{m\rr n_\target n\rr} \nonumber\\
&=\frac{\dot x_{\target,n_\target}\paren{x_{\target,n_{\target}}-x_{\mathrm{tx},m\rr}}+\dot y_{\target,n_\target}\paren{y_{\target,n_{\target}}-y_{\mathrm{rt},m\rr}}}{\lambda r_{m\rr n_\target m}} +\frac{\dot x_{\target,n_\target}\paren{x_{\target,n_{\target}}-x^{\mathrm{rr}}_{n\rr}}+\dot y_{\target,n_\target}\paren{y_{\target,n_{\target}}-y^{\mathrm{rr}}_{n\rr}}}{\lambda r_{n_\target n\rr}}
\label{normalized doppler}
\end{align}
where $b_{\tau,m\rr n\rr}\paren{\cdot}$ and $b_{f,m\rr n\rr}\paren{\cdot}$ are the delay and Doppler observation functions, $\lambda$ and $c$ denote the carrier wavelength and the speed of light, respectively; $r_{\mathrm{r},m\rr n_\target n\rr}$ represents the bistatic range between the $\ith{m\rr}$ radar Tx, $\ith{n}_\target$ target, and $\ith{n\rr}$ radar Rx with
\begin{flalign}
r_{m\rr n_\target m}=&\sqrt{\paren{x_{\target,n_{\target}}-x_{\mathrm{tx},m\rr}}^2+\paren{y_{\target,n_{\target}}-y_{\mathrm{rt},m\rr}}^2}\\
r_{n_\target n\rr}=&\sqrt{\paren{x_{\target,n_{\target}}-x^{\mathrm{rr}}_{n\rr}}^2+\paren{y_{\target,n_{\target}}-y^{\mathrm{rr}}_{n\rr}}^2}.
\end{flalign}
\normalsize
We then write the composite channel coefficient between Tx $m\rr$, target $n_\target$, and Rx $n\rr$ as 
\begin{equation}
    h^{\mathrm{r}}_{m\rr n_\target n\rr}\triangleq r^{-2}_{\mathrm{r},m\rr n_\target n\rr}g_{m\rr n_\target n\rr}\exp\paren{-\mathrm{j}2\pi f_\mathrm{c}\tau_{m\rr n_\target n\rr}},
\end{equation}
\normalsize
where $g_{m\rr n_\target n\rr}$ is the target reflectivity associated with the $\ith{\paren{m\rr,n_{\target},n\rr}}$ path, and $f_\mathrm{c}$ the carrier frequency. 

We also assume that the Swerling I target model holds for each target such that $\mathbf{x}^{\mathrm{t}}_{n_\mathrm{t}}$ and $h^{\mathrm{r}}_{m\rr n_\target n\rr}$  remain constant over a CPI \cite{haimovich2008mimo}.  The inherent nature of a widely distributed MIMO radar determines that its resolution on a range-Doppler plane depends on the target's location and speed and the Txs and Rxs. Therefore, it demands a statistical view of the received radar signal to derive the ambiguity function for a widely distributed MIMO radar. For our multi-target model, we assume that targets are separated by the minimum range-Doppler resolution specified by the statistical ambiguity function of the widely separated MIMO radar \cite{Radmard2014AF}.

Since the RRHs and UEs are also widely distributed, we consider both small-scale fading and distance-dependent path-loss in channel modeling \cite{Lee2013DAS}. Denote the UL channel between the $\ith{i}$ UL UE and the $\ith{m}$ RRH, and the DL channel between the $\ith{m}$ RRH and the $\ith{j}$ DL UE as $\mathbf{h}_{\mathrm{u},im}=r^{-2}_{\mathrm{u},im}\mathbf{g}_{\mathrm{u},im}$ and $\mathbf{h}_{\mathrm{d},mj}=r^{-2}_{\mathrm{d},mj}\mathbf{g}_{\mathrm{d},mj}$, 
where $r_{\mathrm{u},im}$ ($r_{\mathrm{d},mj}$) and $\mathbf{g}_{\mathrm{u},im}\in\mathbb{C}^{M_{\mathrm{c}}}$ ($\mathbf{g}_{\mathrm{d},mj}\in\mathbb{C}^{M_{\mathrm{c}}}$) are the distance and small-scale channel vector between the $\ith{i}$ UL UE and the $\ith{m}$ RRH (the $\ith{m}$ RRH and the $\ith{j}$ DL UE); the path-loss exponent is assumed to be $2$ \cite{Heathvehicularcommradar}. The IBFD transmission and reception at the $\ith{m}$ RRH introduces \textit{self-interfering channel} $\mathbf{H}^{\mathrm{SI}}_{mm}\in\mathbb{C}^{\mathrm{M}\cc\times \mathrm{M}\cc}$ and \textit{inter-RRH} channels $\mathbf{H}^{\mathrm{IR}}_{m^\prime m}\in\mathbb{C}^{\mathrm{M}\cc\times \mathrm{M}\cc}$, for $m^\prime\neq m$. On the other hand, the DL UEs also suffer from co-channel interference due to the UL UEs' transmissions. We model the channel between the $\ith{i}$ UL UE and $\ith{j}$ DL UE as $h_{\mathrm{ud},ij}=r^{-2}_{\mathrm{ud},ij}g_{\mathrm{ud},ij}$.

The D-ISAC system's concurrent transmissions of radar and communications signals imply that the communications and radar signals are overlaid at all Rxs. We model the channels between the $\ith{m}$ RRH/the $\ith{i}$ UL UE and the $\ith{n\rr}$ radar Rx as $\mathbf{h}^{\mathrm{dr}}_{mn\rr}$ ($h^{\mathrm{ur}}_{in\rr}$), and the channel between the $\ith{m\rr}$ radar Tx and the $\ith{m}$ RRH (the $\ith{j}$ DL UE) as $h^{\mathrm{r,u}}_{m\rr,m}$ ($h^{\mathrm{rd}}_{m\rr n_\target j}$).  During the D-ISAC operation window, the RRHs and DL UEs intercept radar signals through direct and target-deflected paths. 

\subsection{Radar signal at the radar Rxs}
\label{sec:RadarSignalRadarRxs}
With the coherent processing and imperfect phase synchronization across the Txs and Rxs, the baseband  received signal model at the $\ith{n\rr}$ radar Rx due to reflections off the $\ith{n_\target}$ is written as
\cite{Arye2010sync,Hongbin2021Sync}
\begin{align}
y^{\mathrm{tr}}_{n_\target n\rr} (t)
&\approx \sum_{m_\mathrm{r}=1}^{M_\mathrm{r}}\sum_{k=0}^{K-1}e^{\mathrm{j}2\pi\bracket{kf'_{m_\mathrm{r}n_{\target}n_\mathrm{r}}+\vartheta_{m\rr n\rr}} } h^{\mathrm{r}}_{m_\mathrm{r}n_{\target}n_\mathrm{r}}a_{k,m\rr}\phi_{m_\mathrm{r}}\paren{t-kT\rr-\tau_{m\rr n_{\target}n\rr}}, \label{approximation} 
\end{align}
\normalsize
where  $f'_{m_\mathrm{r}n_{\target}n_\mathrm{r}}\triangleq f_{m_\mathrm{r}n_{\target}n_\mathrm{r}}T\rr$ is the normalized Doppler frequency and the approximation in (\ref{approximation}) follows the assumption that $f_{m_\mathrm{r}n_{\target} n_\mathrm{r}}\ll\sfrac{1}{T_{\mathrm{r}}}$  \cite{hongbin_movingtarget,MCMIMO_Rad}; $\vartheta_{m\rr n\rr}$ the phase offset between the $\ith{n\rr}$ Rx and $\ith{m\rr}$ Tx.
In practice, synchronization errors can be modeled here as zero-mean Gaussian random
variables with common variance. With sampling rate $F_\mathrm{p}=1/T_\mathrm{p}$, the discrete-time version of $y_{\mathrm{r},n\rr}(t)$ is
\begin{flalign}
&\widetilde{y}^{\mathrm{tr}}_{n_\target n\rr}\bracket{k,n}\triangleq y^\mathrm{tr}_{n_\target n\rr}\paren{kT_{\mathrm{r}}+nT_{\mathrm{p}}}=\sum_{m_\mathrm{r}=1}^{M_\mathrm{r}}e^{\mathrm{j}2\pi kf'_{m_\mathrm{r}n_{\target}n_\mathrm{r}}} h^{\mathrm{r}}_{m\rr n_\target n\rr}a_{k,m\rr}\phi_{m_\mathrm{r}}\bracket{n-n_{m\rr n_\target n\rr}}\nonumber
\end{flalign}
\normalsize
where $\phi_{m_\mathrm{r}}\bracket{n}=\phi_{m\rr}\paren{nT_{\mathrm{p}}}$;
$n_{m\rr n_\target n\rr}=\lfloor \sfrac{\xi_{m\rr n_\target n\rr}}{T_p}\rfloor$ is the discrete delay for the $\ith{\paren{m\rr, n_\target, n\rr}}$ path, which is retrieved through the peak at the output of the $\ith{m\rr}$ matched filter $\phi_{m\rr}\bracket{n}$ at the $\ith{n\rr}$ radar Rx, i.e.,  
\begin{flalign}
y^\mathrm{tr}_{m\rr n_\target n\rr}\bracket{k}\triangleq&\widetilde{y}^\mathrm{tr}_{n_\target n\rr}\bracket{k,n}\ast\phi_{m\rr}^\ast\bracket{n}|_{n=n_{m\rr n_\target n\rr}}\nonumber\\
=&e^{\mathrm{j}2\pi kf^\prime_{m_\mathrm{r}n_{\target}n_\mathrm{r}}} h^{\mathrm{r}}_{m\rr n_\target n\rr}a_{k,m\rr}, \forall k=1,\cdots,K.
\end{flalign}
\normalsize
Combining the $K$ samples yields
\begin{flalign}
\mathbf{y}^\mathrm{tr}_{m\rr n_\target n\rr}=&\bracket{y^\mathrm{tr}_{m\rr n_\target n\rr}\bracket{1},\cdots,y^\mathrm{tr}_{m\rr n_\target n\rr}\bracket{K}}=h^{\mathrm{r}}_{m\rr n_\target n\rr}\mathbf{s}_{m\rr n_\target n\rr},\label{eq:mrnr_radar_Rx}
\end{flalign}\normalsize
where $\mathbf{s}_{m\rr n_\target n\rr}=\mathbf{q}_{m\rr n_\target n\rr}\odot\mathbf{a}_{m\rr}$ and $\mathbf{q}_{m\rr n_\target n\rr}=\bracket{1,\cdots,e^{\mathrm{j}2\pi Kf^\prime_{m_\mathrm{r}n_{\target}n_\mathrm{r}}}}^\top\in\mathbb{C}^{K}$ is the temporal steering vector associated with the $\ith{\paren{m\rr,n_\target,n\rr}}$ radar Tx-target-Rx path.
Applying an $P$-point ($P\geq K$) discrete Fourier transform (DFT) to \eqref{eq:mrnr_radar_Rx} gives the Doppler spectrum of the $\ith{n_{m\rr n_\target,n\rr}}$ range bin 
\begin{flalign}
Y^\mathrm{tr}_{m\rr n_\target n\rr}\bracket{n_{m\rr n_\target,n\rr},p}=&\sum_{k=0}^{K-1}y^\mathrm{tr}_{m\rr n_\target n\rr}\bracket{k}e^{-\mathrm{j}2\pi \frac{kp}{P}}, 
\label{eq:DopplerSpectrum}
\end{flalign}\normalsize
where $\sfrac{p}{P}$ is the $\ith{p}$ normalized Doppler bin for $p = 0,\cdots,P-1$. \eqref{eq:DopplerSpectrum} peaks at the $\ith{\paren{n_{m\rr n_\target,n\rr},p_{m\rr n_\target n\rr}}}$ range-Doppler bin with $p_{m\rr n_\target n\rr}=\lfloor {f^\prime_{m\rr n_\target n\rr}M}\rfloor$. Next we write the measurement vector of the $\ith{n_\target}$ target retrieved by the $\ith{\paren{m\rr,n\rr}}$ radar channel as $\mathbf{z}_{m\rr n_\target n\rr}=\bracket{n_{m\rr n_\target n\rr};p_{m\rr n_\target n\rr}}$. Due to the presence of $N_\target$ targets and $M\rr$ distributed radar Txs, there is uncertainty for the $\ith{n\rr}$ radar Rx to assign a measurement to its corresponding target. In section \ref{sec:estimation}, we employ the JPDA algorithm to ascertain measurements of each target using echoes from all Tx-Rx pairs. 

In practice, apart from the target, the MIMO radar Rxs also receive echoes from undesired targets or clutter, such as buildings and forests. We model the clutter trail at the range cell containing the $\ith{n_\target}$ target at the $\ith{\paren{m\rr,n\rr}}$ radar Tx-Rx pair as $\mathbf{y}^\mathrm{c}_{m\rr n_\target n\rr}=\mathbf{A}\boldsymbol{\rho}_{m\rr n_\target n\rr}\in\mathbb{C}^{K}$, where $\boldsymbol{\rho}_{m\rr n_\target n\rr}\in\mathbb{C}^{M\rr}\sim\mathcal{CN}\paren{0,\sigma^2_{\textrm{c},n\rr}\mathbf{I}_{M\rr}}$ denotes the the clutter component reflection coefficient vector associated with the $\ith{n\rr}$ radar Rx whose covariance matrix (CM) is $\mathbf{R}_{\mathrm{c},n_\target n\rr}=\mathbf{A}\boldsymbol{\Sigma}_{\mathrm{c},n_\target n\rr}\mathbf{A}^\dagger$,  where $\boldsymbol{\Sigma}_{\mathrm{c},n_\target n\rr}=\mathbb{E}\bracket{\boldsymbol{\rho}_{m\rr n_\target n\rr}\boldsymbol{\rho}^\dagger_{n_\target n\rr}}$. 

\subsection{Communications signal at the FD-C-RAN Rxs}
We present the UL and DL signals received by the $\ith{m}$ RRH and $\ith{j}$ DL UE at the $\ith{\bracket{k,l}}$ symbol period as 
\begin{equation}
\mathbf{y}^{\mathrm{uu}}_{m}\bracket{k,l} =\sum_{i=1}^{I} y^\mathrm{u}_{im}\bracket{k,l}=\sum_{i=1}^{I}\mathbf{h}_{\mathrm{u},im}x^\mathrm{u}_{i}\bracket{k,l},
\end{equation}
and
\begin{flalign}
y^{\mathrm{dd}}_{j}\bracket{k,l}=&\mathbf{h}^{\top}_{\mathrm{d},j}\mathbf{x}_{\mathrm{d}}\bracket{k,l}=\sum_{m=1}^{M}\mathbf{h}^{\top}_{\mathrm{d},mj}\mathbf{v}_{\mathrm{d},mj}s_{\mathrm{d},j}\bracket{k,l}+y^{\mathrm{dm}}_{j}
\end{flalign}
\normalsize
where $\mathbf{h}_{\mathrm{d},j}=\bracket{\mathbf{h}^{\top}_{\mathrm{d},1j},\cdots,\mathbf{h}^{\top}_{\mathrm{d},Mj}}^\top\in\mathbb{C}^{MM\cc}$ and $\mathbf{x}_{\mathrm{d}}\bracket{k,l}=\bracket{\mathbf{x}^{\mathrm{d}^\top}_{1}\bracket{k,l},\cdots,\mathbf{x}^{\mathrm{d}^\top}_{M}\bracket{k,l}}^\top$ denote the total DL channel vectors and signals by the $M$ RRHs; $y^{\mathrm{dm}}_{j}=\sum_{m=1}^{M}\mathbf{h}^{\mathrm{d}^\top}_{mj}\sum_{j^\prime\neq j}\mathbf{v}^{\mathrm{d}}_{mj^\prime}s^\mathrm{d}_{j^\prime}\bracket{k,l}$ denotes the DL multi-user interference (MUI) observed by the $\ith{j}$ DL UE. 
\section{FD D-ISAC 
Processing}
\label{sec:fd_proc}
This section presents aspects of processing unique to the FD D-ISAC system. We first introduce the SI and CI models owing to the IBFD transmission. Then we describe the radar signal model observed by communications Rxs and vice versa.  Last, we discuss strategies to handle the synchronization of the distributed system in  
The synchronization 
\subsection{IBFD induced interference}
Since all the RRHs operate in STAR mode, each RRH observes self-interference (SI) and inter-RRH interference (IRI). The SI-plus-IRI arriving at the $\ith{m}$ RRH is  
\begin{flalign}
\mathbf{y}^{\mathrm{SR}}_{m}\bracket{k,l}=&\mathbf{y}^{\mathrm{SI}}_{m}\bracket{k,l}+\mathbf{y}^{\mathrm{IR}}_{m}\bracket{k,l}\nonumber\\
=&\mathbf{H}^{\mathrm{SI}}_{mm}\mathbf{x}^\mathrm{d}_{m}\bracket{k,l}+\sum_{m^\prime\neq m}\mathbf{H}^{\mathrm{IR}}_{m^\prime m}\mathbf{x}^\mathrm{d}_{m^\prime}\bracket{k,l}.
\label{eq:SI-IR}
\end{flalign}
\normalsize
The C-RAN architecture delegates the BBU to process baseband data captured by all the RRHs, which are interconnected via fiber-wired front-haul links. As a result, the BBU has full knowledge of all the DL signals $\braces{x^\mathrm{d}_{m}\bracket{k,l},\forall m,k,l}$ and it subtracts \eqref{eq:SI-IR} upon receiving the data from all RRHs. However, the cancellation is not perfect due to imperfect estimations of $\braces{\mathbf{H}^{\mathrm{SI}}_{m,m^\prime}}$.
Without specifying an SI cancellation technique, we model the residual SI and IRI at the BBU as random Gaussian variables  $\mathbf{y}^{\mathrm{SR}}_{\mathrm{u}}\bracket{k,l}\sim\mathcal{CN}\paren{\mathbf{0},\mathbf{R}^{\mathrm{SR}}_{\mathrm{u}}}$ \cite{Schober16distributedFD}, where 
\begin{equation}
\mathbf{R}^{\mathrm{SR}}_{\mathrm{u}}=\gamma\diag\braces{\sum_{j=1}^{J}\mathbf{H}^{\mathrm{SR}}_{\mathrm{u}}\mathbf{v}_{\mathrm{d},j}\mathbf{v}^{\mathrm{d}^\dagger}_{j}\mathbf{H}^{\mathrm{SR}^\dagger}_{\mathrm{u}}},
\end{equation}\normalsize
$\gamma$ is a constant arising from the output noise of the SR cancellation at the BBU, $\mathbf{H}^{\mathrm{SR}}_{\mathrm{u}}\in\mathbb{C}^{MM\cc\times MM\cc}$ contains $\braces{\mathbf{H}^{\mathrm{SI}}_{mm},\mathbf{H}^{\mathrm{IR}}_{m^\prime m},\forall m}$, and $\mathbf{v}_{\mathrm{d},j}=\bracket{\mathbf{v}^{\mathrm{d}^\top}_{1j},\cdots,\mathbf{v}^{\mathrm{d}^\top}_{Mj}}^\top\in\mathbb{C}^{MM\cc}$. We then show the co-channel interference experienced by the $\ith{j}$ DL UE as 
\begin{flalign}
y^{\mathrm{ud}}_{j}\bracket{k,l}=&\sum_{i=1}^Ih^{\mathrm{ud}}_{ij}\mathbf{x}^\mathrm{u}_{i}\bracket{k,l},
\label{eq:CCI}
\end{flalign}
\normalsize
and its covariance matrix $\mathbf{R}^{\mathrm{ud}}_{j}=\sum_{i=1}^Ih^{\mathrm{ud}}_{ij}\mathbf{v}^{\mathrm{u}^\dagger}_{i}P_{\mathrm{u},i}$ with $P_{\mathrm{u},i}=\bracket{v^{\mathrm{u}}_{i1},\cdots,v^{\mathrm{u}}_{iM}}^\top\in\mathbb{C}^{MM\cc}$.
\subsection{Communications signals received by radar Rxs}
The continuous transmission of the communications signal indicates that all the range bins for each radar Tx-Rx channel contain communications signal components. However, our D-ISAC system design focuses only on the target-occupied range bins. Following the modeling in the first companion paper (Part I) \cite{liu2024codesigningpart1}, we write the DL and UL signals appearing in the range bin of the $\ith{n_\target}$ target at the $\ith{n\rr}$ radar Rx as
\begin{flalign}
y^{\mathrm{dr}}_{m\rr n_\target n\rr}\bracket{k}=&\sum_{m=1}^{M}\mathbf{h}^{\mathrm{dr}}_{ mn\rr}x^\mathrm{d}_{m}\bracket{k,n^{\mathrm{dr},m}_{m\rr n_\target n\rr}}\label{eq:DLradar}\\
y^{\mathrm{ur}}_{m\rr n_\target n\rr}\bracket{k}=&\sum_{i=1}^{I}h^{\mathrm{ur}}_{in\rr}x^\mathrm{u}_{i}\bracket{k,n^{\mathrm{ur},i}_{m\rr n_\target n\rr}}\label{eq:ULradar}
\end{flalign}
\normalsize 
where $n^{\mathrm{dr},m}_{m\rr n_\target n\rr}$ ($n^{\mathrm{ur},i}_{m\rr n_\target n\rr}$) indexes the DL (UL) symbol infringing the range bin where the $\ith{n_\target}$ target is located for the $\ith{\paren{m\rr,n\rr}}$ radar channel.  
Upon obtaining \eqref{eq:DLradar} and \eqref{eq:ULradar}, we next show the composite received signal at the $\ith{\paren{m\rr,n\rr}}$ radar channel regarding the $\ith{n_\target}$ target as
\begin{flalign}
\mathbf{y}^{\mathrm{r}}_{m\rr n_\target n\rr}=\mathbf{y}^{\mathrm{tr}}_{m\rr n_\target n\rr}+\mathbf{y}^{\mathrm{dr}}_{m\rr n_\target n\rr}+\mathbf{y}^{\mathrm{ur}}_{m\rr n_\target n\rr}+\mathbf{y}^{\mathrm{c}}_{m\rr n_\target n\rr}+\boldsymbol{\alpha}^{\mathrm{r}}_{n\rr}
\label{eq: radar_total_received}
\end{flalign}\normalsize
where \eqref{eq:DLradar} and \eqref{eq:ULradar} are the $\ith{k}$ elements of $\mathbf{y}^{\mathrm{dr}}_{m\rr n_\target n\rr}$ and $\mathbf{y}^{\mathrm{ur}}_{m\rr n_\target n\rr}$;
$\alpha^{\mathrm{r}}_{n\rr}\sim\mathcal{CN}\paren{0,\sigma^2_{\mathrm{r},n\rr}\mathbf{I}}$ is the CSCG noise element at the $\ith{n\rr}$ radar Rx with variance $\sigma^2_{\mathrm{r}}$. We also define $\mathbf{y}^{\mathrm{in}}_{\mathrm{r},n_\target n\rr}\triangleq\mathbf{y}^{\mathrm{dr}}_{n_\target n\rr}+\mathbf{y}^{\mathrm{ur}}_{n_\target n\rr}+\mathbf{y}^{\mathrm{c}}_{n_\target n\rr}+\boldsymbol{\alpha}^{\mathrm{r}}_{n\rr}$ as the interference-plus-noise (IN) component of  \eqref{eq: radar_total_received}.

\subsection{Radar signals received by communications users}
The intermittent transmission of the radar Txs results in limited symbol periods being interfered with radar signals. We express the radar signal emitted by the $\ith{m\rr}$ radar Tx arriving at the $\ith{m}$ RRH and $\ith{j}$ DL UE as
\begin{flalign}
\mathbf{y}^{\mathrm{ru}}_{m\rr m}\bracket{k,l}=
\begin{cases}
\mathbf{h}^{\mathrm{\mathrm{ru}}}_{m\rr n_\target m}a_{k,m\rr}, \;&l=l^{\mathrm{u}}_{m\rr n_\target m},\forall m\rr,n_\target,m,\\
\mathbf{0},  &\textrm{elsewise} ,
\end{cases}\label{eq:radar_UL}
\end{flalign}
and
\begin{flalign}
y^{\mathrm{rd}}_{m\rr j}\bracket{k,l}=
\begin{cases}
h^{\mathrm{\mathrm{rd}}}_{m\rr n_\target m}a_{k,m\rr}, \;&l=l^{\mathrm{d}}_{m\rr n_\target j},\forall m\rr,n_\target,j,\\
0, &\textrm{elsewise} ,
\end{cases}\label{eq:radar_DL}
\end{flalign}
\normalsize
where $l^{\mathrm{u}}_{m\rr n_\target m}$ and $l^{\mathrm{d}}_{m\rr n_\target j}$ index the UL/DL symbols interfered by the $\ith{m\rr}$ radar Tx's signal, which are referred to as the symbols of interest for the rest of the paper; $n_\target=0$ and $n_\target=1,\cdots,N_\target$ represent the direct path and the $N_\target$ target-reflection paths, respectively. 

Next we present the comprehensive receive signals at the $\ith{m}$ RRH and the $\ith{j}$ DL UE during the $\ith{\paren{k,l}}$ symbol period as
\begin{flalign}
\label{eq:RRH_symbol}
&\mathbf{y}_{\mathrm{u},m}\bracket{k,l}=\mathbf{y}^{\mathrm{uu}}_{m}\bracket{k,l}+\mathbf{y}^{\mathrm{SR}}_{m}\bracket{k,l}+\sum_{m\rr=1}^{M\rr}\mathbf{y}^{\mathrm{ru}}_{m\rr m}\bracket{k,l}+\boldsymbol{\alpha}_{\mathrm{u},m}
\end{flalign}
and
\begin{flalign}
&y_{\mathrm{d},j}\bracket{k,l}=y^{\mathrm{dd}}_{j}\bracket{k,l}+y^{\mathrm{ud}}_{j}\bracket{k,l}+\sum_{m\rr=1}^{M\rr}y^{\mathrm{rd}}_{m\rr j}\bracket{k,l}+\alpha_{\mathrm{d},j},
\label{eq:DL_symbol}
\end{flalign}
\normalsize
where $\boldsymbol{\alpha}_{\mathrm{u},m}\sim\mathcal{CN}\paren{0,\sigma^2_{\mathrm{u}}\mathbf{I}}$  ($\alpha_{\mathrm{d},j}\sim\mathcal{CN}\paren{0,\sigma^2_{\mathrm{d}}}$) denotes the UL (DL) CSCG  noise element at the $\ith{m}$ RRH ($\ith{j}$ DL UE). The $\ith{m}$ RRH forwards \eqref{eq:RRH_symbol} to the BBU. The composite UL signal collected by the BBU is written as
\begin{flalign}
\mathbf{y}_{\mathrm{u}}\bracket{k,l} =&\bracket{\mathbf{y}^\top_{\mathrm{u},1}\bracket{k,l};\cdots;\mathbf{y}^\top_{\mathrm{u},M}\bracket{k,l}}\nonumber\\ =&\sum_{i=1}^I\mathbf{h}_{\mathrm{u},i}x_{\mathrm{u},i}\bracket{k,l}+\mathbf{y}^{\mathrm{SR}}_{\mathrm{u}}\bracket{k,l}+\sum_{m\rr=1}^{M\rr}\mathbf{y}^{\mathrm{ru}}_{m\rr}\bracket{k,l}+\boldsymbol{\alpha}_{u},
\label{eq:BBU_total}
\end{flalign}\normalsize
where $\mathbf{h}_{\mathrm{u},i}=\bracket{\mathbf{h}^{\top}_{\mathrm{u},i1},\cdots,\mathbf{h}^{\top}_{\mathrm{u},iM}}^\top\in\mathbb{C}^{MM\cc}$, $\boldsymbol{\alpha}_{u}=\bracket{\boldsymbol{\alpha}^\top_{\mathrm{u},1},\cdots,\boldsymbol{\alpha}^{\top}_{\mathrm{u},M}}\in\mathbb{C}^{MM\cc}$ and $\mathbf{y}^{\mathrm{ru}}_{m\rr}\bracket{k,l}=\bracket{\mathbf{y}^{\mathrm{ru}}_{m\rr 1}\bracket{k,l},\cdots,\mathbf{y}^{\mathrm{ru}}_{m\rr M}\bracket{k,l}}\in\mathbb{C}^{MM\cc}$. To decode $s_{\mathrm{u},i}\bracket{k,l}$, the BBU applies the receive beamforming vector $\mathbf{u}_{\mathrm{u},i}\in\mathbb{C}^{MM\cc}$ to \eqref{eq:BBU_total}. The output of $\mathbf{u}_{\mathrm{u},i}\bracket{k,l}$ is given as 
\begin{equation}
\hat{s}_{\mathrm{u},i}\bracket{k,l}=\mathbf{u}^{\dagger}_{\mathrm{u},i}\bracket{k,l}\mathbf{y}_{\mathrm{u}}\bracket{k,l},  \forall i,k,\textrm{and }l.
\end{equation}
\normalsize

\subsection{Synchronization of FD D-ISAC}
\label{sec:syncronization}
Maintaining a desired level of synchronization is an inherently challenging task for distributed systems as due to the presence of multiple channels, timing, and carrier frequency offsets.  The multiple communication
nodes can cooperatively sense the environment, such as in a
C-RAN existing approaches to .....

In \cite{Zhang2022ISACAsync} Estimate carrier frequency offset master-slave paradigm. The carrier and symbol synchronizations are maintained by the FD MU-MIMO communications system by periodically estimating the carrier frequency and phase \cite{Multiuser}. Our proposed FD D-ISAC model requires extensive cooperation between the communications and sensing nodes. The radar Rxs relay the targets' information to the communications Rxs such that $\mathbf{h}^{\mathrm{\mathrm{ru}}}_{m\rr n_\target m}$ ($\mathbf{h}^{\mathrm{\mathrm{rd}}}_{m\rr n_\target m}$) and $l^{\mathrm{u}}_{m\rr n_\target m}$ ($l^{\mathrm{d}}_{m\rr n_\target,j}$) can be estimated by the BBU (the $\ith{j}$ DL UE). Conversely, the BBU feeds the channel information of the DL and UL UEs to the radar Rxs to estimate  $n^{\mathrm{dr},m}_{m\rr n_\target n\rr}$  and $n^{\mathrm{ur},i}_{m\rr n_\target n\rr}$. 
The Rxs of radar and communications employ the same sampling rates; therefore, communications symbols and radar range cells are aligned in time \cite{rihan2018optimum,MCMIMO_RadComm}. The clocks at the BS and the MIMO radar are synchronized offline and periodically updated such that the clock offsets between the BS and MIMO radar Rxs are negligible \cite{interferencealignment}. Using the feedback of the BS via pilot symbols, radar Rxs can obtain the clock information of UL UEs. Note that this setup exploits the established clock synchronization standards that have been widely adopted in wireless communications and distributed sensing systems, e.g., the IEEE 1588 precision time protocol.
\section{FD D-ISAC Design}
\label{sec:solution}
In this section, we discuss the FD D-ISAC system design scheme. We first assign the phases of $\braces{\mathbf{v}_{\mathrm{d},mj}}$ by applying DCP. Subsequently, we derive the DL/UL achievable rates and the MI obtained at each radar Rx for each target to formulate a weighted sum-rate optimization problem w.r.t. the DL/UL beamformers and radar code matrix. Next, we propose an alternating optimization algorithm enabled by the BB low-complexity algorithm.

\subsection{Co-phasing-enabled precoder design}
\label{sec:Co-phasing}
The fundamental of co-phasing is to adjust the phase compensation of each antenna of a
transmitter to match the channel phase such that the signals received from all Txs are constructively superimposed at the Rx. The DCP is an extension of the conventional co-phasing to distributed systems, which enables multiple distributed Txs to coherently transmit a common message signal to a particular client. The architecture of the FD C-RAN designates the BBU to coordinate RRHs and achieve cooperative communications, e.g., Coordinated Multiple Points (CoMP). Therefore, to apply DCP to the DL transmission of the proposed D-ISAC system, 
we recall that the small fading channel vector between the $\ith{m}$ RRH and $\ith{j}$ DL UE $\mathbf{g}_{\mathrm{d},mj}=\bracket{g^{1}_{\mathrm{d},mj},\cdots,g^{M\cc}_{\mathrm{d},mj}}^\top$, where $g^{m\cc}_{\mathrm{d},mj}=\lvert g^{m\cc}_{\mathrm{d},mj}\rvert e^{\mathrm{j}\vartheta^{m\cc}_{\mathrm{d},mj}}$ and $\vartheta^{m\cc}_{\mathrm{d},mj}$ are the channel coefficient and phase associated with the $\ith{m\cc}$ antenna at the $\ith{m}$ RRH, respectively.   As a prerequisite for DCP, we suppose that $\braces{\vartheta^{m\cc}_{\mathrm{d},m\cc j}}_{m\cc,j=1}^{M\cc,J}$ are estimated using known pilot symbols via methods such as maximum likelihood \cite{chopra2016multistream}, and are available at the $\ith{m}$ RRH, we express DCP-enabled $\mathbf{v}_{\mathrm{d},mj}$ as \cite{Lee2013DAS,Liu2022DCP}
\begin{equation} 
\label{cophasingweights}
\mathbf{v}_{\mathrm{d},mj} =\sqrt{P_{\mathrm{d},mj}}\frac{\mathbf{g}^{\dagger}_{\mathrm{d},mj}}{\lVert\mathbf{g}_{\mathrm{d},mj}\rVert},
\end{equation}
\normalsize
where $P_{\mathrm{d},mj}$ denotes the power allocated to the $\ith{j}$ DL UE by the $\ith{m}$ RRH, which will be optimized in the next section. The total DL transmit power by the $\ith{m}$ RRH is $P_{\mathrm{d}}=\sum_{j=1}^JP_{\mathrm{d},mj}$ for all $m=1,\cdots,M$. We also define the super DL beamformer for the $\ith{j}$ DL UE as $\mathbf{v}_{\mathrm{d},j}=\bracket{\mathbf{v}^\top_{\mathrm{d},mj},\cdots,\mathbf{v}^\top_{\mathrm{d},Mj}}^\top\in\mathbb{C}^{MM\cc}$
\subsection{Weighted sum-rate maximization}
\label{subsec: formulation}
The performance metrics for designing radar and communications systems are not identical because of different system goals. For example, a communications system generally strives to achieve high data rates, while a radar performs detection, estimation, and tracking. Some recent works \cite{alaee2020information,dokhanchi2020multi} suggest MI as a common performance metric. The MI is a well-studied metric in MU-MIMO communications for transmitting precoder design \cite{Luo2011IterativeWMMSE}. The seminal work on radar waveform design metric by \cite{Bellinformation} originally proposed MI as a measure of radar performance. Later, MI-based waveform design was also extended to MIMO radars \cite{Jammer_game, NaghshTSP2017}. It has been shown \cite{Jammer_game} that maximizing the MI between the radar received signal and the target response leads to a better detection performance in the presence of the Gaussian noise.

In this section, we extend the information-theoretic performance metric proposed in the first companion paper (Part I) \cite{liu2024codesigningpart1} to the D-ISAC system design with the presence of multiple targets. The major difference is that the performance metric in the first companion paper (Part I) \cite{liu2024codesigningpart1} is based on a single range cell, a.k.a., cell-under-test, and herein we expand it to the entire range profile in light of the presence of multiple targets. 

To derive the information-theoretic performance metric for the D-ISAC system, we express the achievable rates of the $\ith{i}$ UL UE and $\ith{j}$ DL UE during the symbols of interest as 
$\mathit{R}^{\mathrm{u}}_{i}\bracket{k,l^{\mathrm{u}}_{m\rr n_\target m}}=\log_2\paren{1+\varsigma^{\mathrm{u}}_{i}\bracket{k,l^{\mathrm{u}}_{m\rr n_\target n\rr}}}$ and $\mathit{R}^{\mathrm{d}}_{j}\bracket{k,l^{\mathrm{d}}_{m\rr n_\target j}}=\log_2\paren{1+\varsigma^{\mathrm{d}}_{j}\bracket{k,l^{\mathrm{d}}_{m\rr n_\target j}}}$, where 
\begin{align}
\label{UL_rate}
\varsigma^{\mathrm{u}}_{i}\bracket{k,l^{\mathrm{u}}_{m\rr n_\target n\rr}}&=\frac{\lvert\mathbf{u}^\dagger_{\mathrm{u},i}\mathbf{h}_{\mathrm{u},i}\rvert^2}{\sum_{q\neq i}\lvert \mathbf{g}^{\dagger}_{\mathrm{u},q}\mathbf{h}_{\mathrm{u},q}\rvert^2+\trace\braces{\mathbf{R}^{\mathrm{SR}}_{\mathrm{u}}}+\sum_{m=1}^{M}\lvert h^{\mathrm{ru}}_{m\rr n_\target m}a_{k,m\rr}\rvert^2+\sigma^2_{\mathrm{u}}}
\end{align}
 and 
\begin{align}
\label{DL_rate}
\varsigma^{\mathrm{d}}_{j}\bracket{k,l^{\mathrm{d}}_{m\rr n_\target j}}&=\frac{\lvert \mathbf{h}^{\dagger}_{\mathrm{d},j}\mathbf{v}_{\mathrm{d},j}\rvert^2}{\sum_{j^\prime\neq j}\lvert \mathbf{h}^{\dagger}_{\mathrm{d},j}\mathbf{v}_{\mathrm{d},j^\prime}\rvert^2+\trace\braces{\mathbf{R}^{\mathrm{ud}}_{j}}+\lvert h^{\mathrm{rd}}_{m\rr n_\target j}a_{k,m\rr}\rvert^2+\sigma^2_{\mathrm{d}}}
\end{align}
are the signal-to-interference-plus-noise-ratios (SINRs) at the $\ith{m}$ RRH and the $\ith{j}$ DL UE, respectively. The information-theoretic metric for the MIMO radar is expressed as the MI between $\mathbf{y}^{\mathrm{r}}_{m\rr n_\target n\rr}$ and $h_{m\rr n_\target n\rr}$ \cite{Liu2022precoder}, i.e., 
\begin{align}
R^{\mathrm{r}}_{m\rr n_\target n\rr}&\triangleq\mathit{I}\paren{\mathbf{y}^{\mathrm{r}}_{m\rr n_\target n\rr};h_{m\rr n_\target n\rr}|\mathbf{a}_{m\rr}}\nonumber\\
&=\log\det\paren{\mathbf{I}_K+\sigma^2_{m\rr n_\target n\rr}\mathbf{s}_{m\rr n_\target n\rr}\mathbf{s}^\dagger_{m\rr n_\target n\rr}\mathbf{R}^{\mathrm{in}^{-1}}_{m\rr n_\target n\rr}},
\end{align}
where $\mathbf{R}^{\mathrm{in}}_{m\rr n_\target n\rr}=\mathbb{E}\bracket{\mathbf{y}^{\mathrm{in}}_{m\rr n_\target n\rr}\paren{\mathbf{y}^{\mathrm{in}}_{m\rr n_\target n\rr}}^\dagger}$ and $\sigma_{m\rr n_\target n\rr}$ is the variance of $h_{m\rr n_\target n\rr}$.
Thus the compounded weighted sum-rate (CWSR) for the D-ISAC is 
\begin{align}
\label{eq: WSR}
R_{\mathrm{CWSR}}&=\sum_{m\rr=1}^{M\rr}\sum_{n\rr=1}^{N\rr}\sum_{n_\target=1}^{N_\target}\alpha^{\mathrm{r}}_{m\rr 
n_\target n\rr}R^{\mathrm{r}}_{m\rr n_\target n\rr}+
\sum_{k=1}^{K}\sum_{m\rr=1}^{M\rr}\sum_{n_\target=1}^{N_\target}\bracket{\sum_{i=1}^{I}\alpha^\textrm{u}_i\mathit{R}^{\mathrm{u}}_{i}\bracket{k,l^{\mathrm{u}}_{m\rr n_\target m}}+\sum_{j=1}^\mathit{J}\alpha^\textrm{d}_j\mathit{R}^{\mathrm{d}}_{j}\bracket{k,l^{\mathrm{d}}_{m\rr n_\target j}}},
\end{align}
where $\alpha^{\mathrm{r}}_{m\rr n_\target n\rr}$, $\alpha^\mathrm{u}_{i}$, and $\alpha^\textrm{d}_j$ are pre-defined weights assigned to the  $\ith{\paren{m\rr},n\rr}$ radar Tx-Rx path, $\ith{i}$ UL and $\ith{j}$ DL UEs, respectively. 

As we have designed the phase terms of $\mathbf{v}_{\mathrm{d},mj}$ in Section \ref{sec:Co-phasing}, our goal is shifted to jointly optimize the radar code matrix $\mathbf{A}$, UL beamformers $\braces{\mathbf{u}_{\mathrm{u},i}}_{i=1}^{I}$, the DL powers $\braces{P_{\mathrm{d},mj}}_{m,j=1}^{M,J}$ and the UL powers $\braces{P_{\mathrm{u},i}}_{i=1}^I$ by maximizing \eqref{eq: WSR} given the DL/UL power budget and MIMO radar waveform constraints,  
\begin{subequations}\label{eq:opt1}
\begin{align}
\underset{\braces{P_{\mathrm{d},mj},P_{\mathrm{u},i},\mathbf{u}_{\mathrm{u},i}, \forall i,j,m},\mathbf{A}}{\text{maximize}}&\quad R_{\mathrm{CWSR}} \label{eq: op1_cost_function}\\
\text{subject to}&\quad
P_{\mathrm{u},i}\leq P_{\mathrm{u},\max},\forall i, \label{constraint:UL_power}\\
&\sum_{j=1}^{J}P_{\mathrm{d},mj}\leq P_{\mathrm{d},\max},\forall  m, \label{constraint:DL_power}\\ 
 & \lVert\mathbf{a}_{m\rr}\rVert^2 =P_{\textrm{r},m\rr},\; \label{constraint:radarpower}\\*
&\frac{\mathrm{\mathit{K}}\max_{k=1,\cdots,  K}\lvert\mathbf{a}_{m\rr}\bracket{k}\rvert^2}{P_{\mathrm{r},m\rr}}\leq\mathrm{\gamma}_{m\rr},\; \forall\; m\rr,\label{constraint:radarpar}
\end{align}
\end{subequations}
where $P_{\mathrm{d},\max}$ and $P_{\mathrm{u},\max}$ are the DL and UL power budgets, \eqref{constraint:radarpower} and \eqref{constraint:radarpar} enforce the PAR constraint for the distributed MIMO radar waveform. \eqref{eq:opt1} is known to be an NP-hard non-convex problem as \eqref{eq: op1_cost_function} is not concave over variables to be optimized jointly, and \eqref{constraint:radarpower} is a non-convex constraint, which makes the global optimum of \eqref{eq:opt1} unobtainable in polynomial time \cite{Liu2018Gloabalsip}. We partition \eqref{eq:opt1} into two subproblems as follows
\begin{subequations}\label{eq:subop1}
\begin{align}
\underset{\braces{P_{\mathrm{d},mj},P_{\mathrm{u},i},\mathbf{u}_{\mathrm{u},i}, \forall i,j,m},\mathbf{A}^\prime}{\text{maximize}}&\quad R_{\mathrm{CWSR}} \\
	\text{subject to}&\;
	\eqref{constraint:UL_power}, \eqref{constraint:DL_power} \nonumber\\ 
	&\lVert\mathbf{a}_{m\rr}\rVert^2\leq P_{\mathrm{r},m\rr},\;\forall k,\label{constraint:radar_power_sub1}
\end{align}
\end{subequations}
and
\begin{align}
\label{eq:subop2}
\underset{\mathbf{a}_{m\rr},\forall m\rr}{\text{minimize}}\;&\lVert\mathbf{a}_{m\rr}-\mathbf{a}^\prime_{m\rr}\rVert^2_2\nonumber\\
\text{ subject to}\; &\eqref{constraint:radarpower} \textrm{ and } \eqref{constraint:radarpar}
\end{align}\normalsize
where $P_{\mathrm{r},\max}=KP_{\mathrm{r},m\rr}$ and $\mathbf{A}^\prime=\bracket{\mathbf{a}^\prime_{1},\cdots,\mathbf{a}^\prime_{M\rr}}$ is the intermediate solution to the radar code matrix from \eqref{eq:subop1}. 
\subsection{Low-complexity solution to Problem \texorpdfstring{\eqref{eq:subop1}}{Eq}}
\label{subsec:solution}
To combat the non-convexity of \eqref{eq:subop1}, we utilize the equivalence between maximizing the achievable rate and minimizing the weighted minimum-mean-square-error (WMMSE) and map \eqref{eq:subop1} to a WMMSE minimization problem as explained in the second companion paper (Part II) \cite{liu2024codesigningpart2}. Denote the MSEs associated with the $\ith{\paren{m\rr n_\target n\rr}}$ radar path,  $\ith{\paren{k,l}}$ UL and DL symbol periods as
\begin{flalign}
E_{\mathrm{r},m\rr n_\target n\rr}
=&\sigma^2_{m\rr n_\target n\rr}-2\sigma^2_{m\rr n_\target n\rr}\mathbf{s}^\dagger_{m\rr n_\target n\rr}\mathbf{u}_{\mathrm{r},m\rr n_\target n\rr}\nonumber\\
&+\mathbf{u}^\dagger
_{\mathrm{r},m\rr n_\target n\rr}\mathbf{R}_{m\rr n_\target n\rr}\mathbf{u}_{\mathrm{r},m\rr n_\target n\rr}, \label{eq:radar_MSE}
\end{flalign}
\begin{flalign}
E_{\mathrm{u},i}\bracket{k,l}=1-2\sqrt{P_{\mathrm{u},i}}\mathbf{h}^\dagger_{\mathrm{u}}\mathbf{u}_{\mathrm{u},i}\bracket{k,l}+\mathbf{u}^\dagger_{\mathrm{u},i}\bracket{k,l}\mathbf{R}_{\mathrm{u}}\bracket{k,l}\mathbf{u}_{\mathrm{u},i}\bracket{k,l},\label{eq:UL_MSE}
\end{flalign}
and
\begin{flalign}
E_{\mathrm{d},j}\bracket{k,l}=&1-2\sum_{m=1}^{M}u^\ast_{\mathrm{d},j}\bracket{k,l}\mathbf{h}^\top_{\mathrm{d},mj}\mathbf{v}_{\mathrm{d},mj}+\frac{u^\ast_{\mathrm{d},j}\bracket{k,l}u_{\mathrm{d},j}\bracket{k,l}}{\sigma^{2}_{\mathrm{d},j}\bracket{k,l}} ,\label{eq:DL_MSE}
\end{flalign}
\normalsize
where $\mathbf{u}_{\mathrm{r},m\rr n_\target n\rr}\in\mathbb{C}^{K}$ and $u_{\mathrm{d},j}\bracket{k,l}$ are the receive filters at the $\ith{\paren{m\rr n_\target n\rr}}$ radar path and the $\ith{j}$ DL UE. To formulate the WMMSE minimization problem, we introduce auxiliary weight variables associated with $E_{\mathrm{r},m\rr n_\target n\rr}$, $E_{\mathrm{u},i}\bracket{k,l}$, and $E_{\mathrm{d},j}\bracket{k,l}$ as $W_{\mathrm{r},m\rr n_\target n\rr}$, $W_{\mathrm{u},i}\bracket{k,l}$, and $W_{\mathrm{d},j}\bracket{k,l}$. Define
\begin{align}
\Gamma&\triangleq\sum_{n\rr=1}^{N\rr}\sum_{m\rr=1}^{M\rr}\sum_{n_\target=1}^{N_\target}\alpha^{\mathrm{r}}_{m\rr n_\target n\rr}W_{\mathrm{r},m\rr n_\target n\rr}E_{\mathrm{r},m\rr n_\target n\rr}  +\sum_{k=1}^{K}\sum_{m\rr=1}^{M\rr}\sum_{n_\target=1}^{N_\target}\sum_{i=1}^{I}\alpha^\mathrm{u}_{i}W_{\mathrm{u},i}\bracket{k,l^{\mathrm{u}}_{m\rr n_\target m}}E_{\mathrm{u},i}\bracket{k,l^{\mathrm{u}}_{m\rr n_\target m}} \nonumber\\
&+\sum_{k=1}^{K}\sum_{m\rr=1}^{M\rr}\sum_{n_\target=1}^{N_\target}\sum_{j=1}^\mathit{J}\alpha^\textrm{d}_jW_{\mathrm{d},j}\bracket{k,l^{\mathrm{d}}_{m\rr n_\target j}}E_{\mathrm{d},j}\bracket{k,l^{\mathrm{d}}_{m\rr n_\target j}}.
\end{align}
Then, the optimization problem becomes
\begin{align}
\underset{\mathbb{W},\mathbb{U},\mathbb{P},\mathbf{A}^\prime}{\textrm{minimize}}&\quad\Gamma\label{eq:WMMSE}\\
\textrm{subject to}&\quad \eqref{constraint:UL_power}, \eqref{constraint:DL_power}  \text{ and } \eqref{constraint:radar_power_sub1}, \nonumber
\end{align}
where $\mathbb{W}\triangleq\braces{W_{\mathrm{r},m\rr n_\target n\rr},W_{\mathrm{u},i},W_{\mathrm{d},j},\forall m\rr,n_\target,n\rr,i,j}$, $\mathbb{U}\triangleq\braces{\mathbf{u}_{\mathrm{r},m\rr n_\target n\rr},\mathbf{u}_{\mathrm{u},i},u_{\mathrm{d},j},\forall m\rr,n_\target,n\rr,i,j}$, and $\mathbb{P}\triangleq\braces{P_{\mathrm{u},i},P_{\mathrm{d},mj},\forall m,j}$. The second companion paper (Part II) \cite{liu2024codesigningpart2} shows that \eqref{eq:WMMSE} yields the same solution as \eqref{eq:subop1}. Next, we solve \eqref{eq:WMMSE} sequentially with the BCD algorithm and update each variable in a Gauss-Seidel manner. We also proved in the second companion paper (Part II) \cite{liu2024codesigningpart2}  that WMMSE solution is proved to be optimal for $\mathbb{U}$. We write $\braces{\mathbf{u}^\star_{\mathrm{u},i}}$ as
\begin{flalign}
\mathbf{u}^\star_{\mathrm{u},i}\bracket{k,l}=&\arg\min_{\mathbf{u}_{\mathrm{u},i}\bracket{k,l},\forall i,k,l}W_{\mathrm{u},i}\bracket{k,l}E_{\mathrm{u},i}\bracket{k,l}\nonumber\\
&=\sqrt{P_{\mathrm{u},i}}\mathbf{R}^{-1}_{\mathrm{u}}\bracket{k,l}\mathbf{h}_{\mathrm{u},i},
\end{flalign}
\normalsize
where $\mathbf{R}_{\mathrm{u}}\bracket{k,l}=\mathbb{E}\bracket{\mathbf{y}_\mathrm{u}\bracket{k,l}\mathbf{y}^\dagger_\mathrm{u}\bracket{k,l}}$. Similarly, we find $\mathbf{u}^\star_{m\rr n_\target n\rr}=\sigma^2_{m\rr n_\target n\rr}\mathbf{R}^{-1}_{m\rr n_\target n\rr}\mathbf{s}_{m\rr n_\target n\rr}$ and $u^\star_{\mathrm{d},j}\bracket{k,l}=\sfrac{\sum_{m=1}^M\mathbf{h}^\top_{\mathrm{d},mj}\mathbf{v}_{\mathrm{d},mj}}{\sigma^{2}_{\mathrm{d},j}\bracket{k,l}}$, which are substituted in \eqref{eq:radar_MSE}-\eqref{eq:DL_MSE} to yield the optimal MSEs $E^\star_{\mathrm{r},m\rr n_\target n\rr}$, $E^\star_{\mathrm{u},i}\bracket{k,l}$, and $E^\star_{\mathrm{d},j}\bracket{k,l}$. The optimal weights are given as $W^\star_{\mathrm{r},m\rr n_\target n\rr}=\paren{E_{\mathrm{r},m\rr n_\target n\rr}}^{-1}$, $W^\star_{\mathrm{u},i}\bracket{k,l}=\paren{E_{\mathrm{u},i}\bracket{k,l}}^{-1}$, and $W^\star_{\mathrm{d},j}\bracket{k,l}=\paren{E_{\mathrm{d},j}\bracket{k,l}}^{-1}$. Given $\mathbb{W^\star}$ and $\mathbb{U^\star}$, we obtain 
\begin{flalign}
\underset{\mathbb{P},\mathbf{A}^\prime}{\textrm{minimize}}\quad\Gamma\paren{\mathbb{W^\star},\mathbb{U^\star}}\;\textrm{subject to}\quad \eqref{constraint:UL_power}, \eqref{constraint:DL_power}  \text{ and } \eqref{constraint:radar_power_sub1}, \label{eq:WMMSE2}
\end{flalign}
which is multi-convex and holds the strong duality condition for one variable when the rest is fixed. This enables us to solve \eqref{eq:WMMSE2} through a Lagrange dual method, as shown in the second companion paper (Part II) \cite{liu2024codesigningpart2}.

Denote the Lagrange multiplier vectors for \eqref{constraint:UL_power}, \eqref{constraint:DL_power}, and \eqref{constraint:radar_power_sub1} by $\boldsymbol{\lambda}_{\mathrm{u}}=\bracket{\lambda_{\mathrm{u},1},\cdots,\lambda_{\mathrm{u},I}}^\top\in\mathbb{R}^{I}$, $\boldsymbol{\lambda}_{\mathrm{d}}=\bracket{\lambda_{\mathrm{d},1},\cdots,\lambda_{\mathrm{d},M}}^\top\in\mathbb{R}^{M}$, and  $\boldsymbol{\lambda}_{\mathrm{r}}\triangleq\bracket{\lambda_{\mathrm{r},1},\cdots,\lambda_{\mathrm{r},m\rr}}^\top\in\mathbb{R}^{K}$,  respectively,
and formulate $\mathbf{p}_{\mathrm{u}}\in\mathbb{R}^{I}$, $\mathbf{p}_{\mathrm{d}}\in\mathbb{R}^{M}$, and $\mathbf{p}_{\mathrm{r}}\in\mathbb{R}^{M\rr}$, where the $\ith{i}$ element of $\mathbf{p}_{\mathrm{u}}$ is $P_{\mathrm{u},i}$, the $\ith{m}$ element of $\mathbf{p}_{\mathrm{d}}$ is $\sum_{j=1}^JP^\mathrm{d}_{mj}$, and the $\ith{m\rr}$ elements of $\mathbf{p}_{\mathrm{r}}$ is  $\lVert\mathbf{a}_{m\rr}\rVert^2$, respectively. We also define $\braces{\boldsymbol{\lambda}}\triangleq\braces{\boldsymbol{\lambda}_{\mathrm{u}},\boldsymbol{\lambda}_{\mathrm{d}},\boldsymbol{\lambda}_{\mathrm{r}}}$,  leading to the Lagrangian associated with \eqref{eq:WMMSE2} as  with the 
\begin{flalign}
\label{Lagrange}
&\mathcal{L}\paren{\braces{P_{\mathrm{d},mj}},\braces{P_{\mathrm{u},i}},\mathbf{A},\braces{\boldsymbol{\lambda}}}=\Gamma+\boldsymbol{\lambda}^\top_{\mathrm{u}}\paren{\mathbf{p}_{\mathrm{u}}-P_{\mathrm{u},\max}\mathbf{1}_{I}}\nonumber\\
&+\boldsymbol{\lambda}^\top_{\mathrm{d}}\paren{\mathbf{p}_{\mathrm{d}}-P_{\mathrm{d},\max}\mathbf{1}_{M}}+ \boldsymbol{\lambda}^\top_{\mathrm{r}}\paren{\mathbf{p}_{\mathrm{r}}-P_\mathrm{r}\mathbf{1}_{K}}.
\end{flalign}\normalsize 
Invoking the Karush-Kuhn-Tucker (KKT) conditions yields
\begin{flalign}\label{eq:dPiB}
&\nabla_{P_{\mathrm{u},i}}\mathcal{L}=\nabla_{\PiB}\Gamma+\lambda_{\mathrm{u},i}P_{\mathrm{u},i}=0,
\end{flalign}
\begin{flalign}\label{eq:dPBj}
&\nabla_{\PBj}\mathcal{L}=\nabla_{\PBj}\Gamma+\lambda_{\mathrm{d},m}\PBj=0,
\end{flalign}
\begin{flalign}\label{eq:damr}
&\nabla_{\mathbf{a}^\prime_{m\rr}}\mathcal{L}=\nabla_{\mathbf{a}^\prime_{m\rr}}\Gamma+\lambda_{\mathrm{r},m\rr}P_{\mathrm{r},m\rr}=0.
\end{flalign}
\normalsize
In order to obtain the closed-form solutions to $P_{\mathrm{u},i}$, $P_{\mathrm{d},j}$, and $\mathbf{a}^\prime_{m\rr}$ through  \eqref{eq:dPiB}-\eqref{eq:damr}, we need to solve the Lagrange dual problem
\begin{flalign}
\label{dualproblem}
&\text{maximize} \quad D\paren{\braces{\boldsymbol{\lambda}}}=\underset{{\braces{P_{\mathrm{d},mj}},\braces{P_{\mathrm{u},i}},\mathbf{A}}}\inf \mathcal{L}\paren{\braces{P_{\mathrm{d},mj}},\braces{P_{\mathrm{u},i}},\mathbf{A},\braces{\boldsymbol{\lambda}}}\nonumber\\
&\text{subject to}\quad  \boldsymbol{\lambda}_{\mathrm{u}},\boldsymbol{\lambda}_{\mathrm{d}},\boldsymbol{\lambda}_{\mathrm{r}} \succeq 0.
\end{flalign}\normalsize

We proceed to solve $\paren{\ref{dualproblem}}$ utilizing the subgradient method to determine  $\boldsymbol{\lambda}^\star_\mathrm{u}$, $\boldsymbol{\lambda}^\star_\mathrm{d}$, and $\boldsymbol{\lambda}^\star_\mathrm{r}$. Using a gradient-descent type optimization method to update $\lambda_{\mathrm{u},i}$, $\lambda_{\mathrm{d},j}$ and $\lambda_{\mathrm{r},m\rr}$ in the $\ith{t}$ iteration yields \cite{BBSUmmwave} 
\begin{flalign}
\label{lambda_UL}
\lambda^{\paren{t+1}}_{\mathrm{u},i}=&\bracket{\lambda^{\paren{t}}_{\mathrm{u},i}+\beta^{\paren{t}}_{\mathrm{u},i}\nabla D\paren{\lambda^{\paren{t}}_{\mathrm{u},i}}}^+\\ 
\lambda^{\paren{t+1}}_{\mathrm{d},m}=&\bracket{\lambda^{\paren{t}}_{\mathrm{d},m}+\beta^{\paren{t}}_{\mathrm{d},m}\nabla D\paren{\lambda^{\paren{t}}_{\mathrm{d},j}}}^+,\label{lambda_DL}\\
\lambda^{\paren{t+1}}_{\mathrm{r},m\rr} =&\bracket{\lambda^{\paren{t}}_{\mathrm{r},m\rr}+\beta^{\paren{t}}_{\mathrm{r},m\rr}\nabla D\paren{\lambda^{\paren{t}}_{\mathrm{r},m\rr}}}^+\label{eq:lamnda_radar}
\end{flalign}
\normalsize
where  $P^{\paren{t}}_{\mathrm{d}}=\sum_{j=1}^{J}P^{\paren{t}}_{\mathrm{d},mj}$; $\beta^{\paren{t}}_{\mathrm{u},i}$, $\beta^{\paren{t}}_{\mathrm{d},k}$, and $\beta^{\paren{t}}_{\mathrm{r},m\rr}$  are the BB step-sizes in the $\ith{t}$ iterations; $\lambda^{\paren{t}}_{\mathrm{u},i}$, $\lambda^{\paren{t}}_{\mathrm{d},m}$, $\lambda^{\paren{t}}_{\mathrm{r},m\rr}$ are the $\ith{t}$ iterates of $\lambda_{\mathrm{u},i}$, $\lambda_{\mathrm{d},m}$, and $\lambda_{\mathrm{r},m\rr}$ with $\nabla D\paren{\lambda^{\paren{t}}_{\mathrm{u},i}}=P^{\paren{t}}_{\mathrm{u},i}-P_{\mathrm{u},\max}$, $\nabla D\paren{\lambda^{\paren{t}}_{\mathrm{d},j}}=P^{\paren{t}}_{\mathrm{d}}-P_{\mathrm{d},\max}$, and $\nabla D\paren{\lambda^{\paren{t}}_{\mathrm{r},m\rr}}=\lVert\mathbf{a}^{\paren{t}}_{m\rr}\rVert^2-P_{\mathrm{r},\max}$ their corresponding sub-gradients.

There are various methods to determine the step-size. The most basic solution is known as line search or backtracking. This strategy reduces the step length in each iteration until Armijo's condition is satisfied, which involves the evaluation of the cost function and its derivative at each iteration. This increases the per-iterate complexity. Polyak's step-size rule, on the other hand, achieves faster descent by using the current gradient to estimate the line search geometry but, as mentioned in the second companion paper (Part II) \cite{liu2024codesigningpart2}, it requires estimating the optimal value of the cost function. However, these methods only employ the gradient and disregard the Hessian of the cost function. We propose to find the step-size with the BB approach, which embeds the second-order information in the step length calculation without computing the Hessian directly. Therefore, the BB approach not only delivers increased performance but also preserves the simplicity of the gradient-type algorithms. We present $\beta^{\paren{t}}_{\mathrm{u},i}$ according to the BB method as follows \cite{BBSUmmwave}:
\begin{equation}
\label{eq: BBstep}
\beta^{\paren{t}}_{\mathrm{u},i}=\frac{\lVert s^{\paren{t}}_{\mathrm{u},i}\rVert^2}{s^{\paren{t}}_{\mathrm{u},i}\gamma^{\paren{t}}_{\mathrm{u},i}},\;\beta^{\paren{t}}_{\mathrm{d},m}=\frac{\lVert s^{\paren{t}}_{\mathrm{d},m}\rVert^2}{s^{\paren{t}}_{\mathrm{d},m}\gamma^{\paren{t}}_{\mathrm{d},m}},\;\beta^{\paren{t}}_{\mathrm{r},m\rr}=\frac{\lVert s^{\paren{t}}_{\mathrm{r},m\rr}\rVert^2}{s^{\paren{t}}_{\mathrm{r},m\rr}\gamma^{\paren{t}}_{\mathrm{r},m\rr}},
\end{equation}\normalsize    
where $s^{\paren{t}}_{\mathrm{u},i}=\lambda^{\paren{t}}_{\mathrm{u},i}-\lambda^{\paren{t-1}}_{\mathrm{u},i}$, $s^{\paren{t}}_{\mathrm{d},m}=\lambda^{\paren{t}}_{\mathrm{d},m}-\lambda^{\paren{t-1}}_{\mathrm{d},m}$, $s^{\paren{t}}_{\mathrm{r},m\rr}=\lambda^{\paren{t}}_{\mathrm{r},m\rr}-\lambda^{\paren{t-1}}_{\mathrm{r},m\rr}$, $\gamma^{\paren{t}}_{\mathrm{u},i}=D\paren{\lambda^{\paren{t}}_{\mathrm{u},i}}-D\paren{\lambda^{\paren{t-1}}_{\mathrm{u},i}}$, $\gamma^{\paren{t}}_{\mathrm{d},m}=D\paren{\lambda^{\paren{t}}_{\mathrm{d},m}}-D\paren{\lambda^{\paren{t-1}}_{\mathrm{d},m}}$, and $\gamma^{\paren{t}}_{\mathrm{r},m\rr}=D\paren{\lambda^{\paren{t}}_{\mathrm{r},m\rr}}-D\paren{\lambda^{\paren{t-1}}_{\mathrm{r},m\rr}}$.  We summarize using the BB algorithm to find $P_{\mathrm{u},i}$ in Algorithm \ref{BBalgorithm}, where $t_{\max}$ is the maximum number of iterations for the BB method and  $\Gamma_{\min}$ tracks the minimum of the cost function. $P^{\star
}_{\mathrm{d},mj}$ and $\mathbf{a}^\prime_{m\rr}$ are found following the same procedure.
\begin{algorithm}[ht!]
\caption{BB Algorithm to Solve $P_{\mathrm{u},i}$}
\label{BBalgorithm}
\begin{algorithmic}[1]
\Statex \textbf{Input:} $t_{\textrm{max}}$, $P_{\mathrm{u},\max}$ 
\Statex \textbf{Output: } $P_{\mathrm{u},i}$ 
\State Set the iteration index $\mathrm{t=0}$; initialize $\lambda^{\paren{0}}_{\mathrm{u},i}$, $D\paren{\lambda^{\paren{0}}_{\mathrm{u},i}}$, $\beta^{\paren{0}}_{\mathrm{u},i}$, $\Gamma^{\paren{0}}$
\State Set $\Gamma_{\min}=\Gamma^{\paren{0}}$
\Repeat
\State Update $\lambda^{\paren{t+1}}_{\mathrm{u},i}$ with \eqref{lambda_UL}
\State Update $\beta^{\paren{t+1}}_{\mathrm{u},i}$ with \eqref{eq: BBstep}
\State Substitute $\beta^{\paren{t+1}}$ in \eqref{eq:dPiB} to find $P^{\paren{t+1}}_{\mathrm{u},i}$ 
\State Update $\Gamma^{\paren{t+1}}$ with $P^{\paren{t+1}}_{\mathrm{u},i}$ 
\If{$\Gamma^{\paren{t+1}}<\Gamma_{\min}$} $\Gamma_{\min}=\Gamma^{\paren{t+1}}$, $P^\star_{\mathrm{u},i}=P^{\paren{t+1}}_{\mathrm{u},i}$ \EndIf
\State	$\mathrm{t}\leftarrow \mathrm{t}+1$ 
\Until $\mathrm{\iota}>\mathrm{\iota}_{\textrm{max}}$
\State \Return $P^\star_{\mathrm{u},i}$
\end{algorithmic}
\end{algorithm}
We then iterate across all variables through an iterative BCD descent approach. To impose the PAR constraint on $\mathbf{a}^\prime_{m\rr}$ via \eqref{eq:subop2}, we resort to a tight-frame based nearest vector method described in Algorithm 3 of the second companion paper (Part II) \cite{liu2024codesigningpart2}. Finally, we summarize the alternating algorithm to solve \eqref{eq:subop1} and \eqref{eq:subop2} sequentially in Algorithm \ref{convexalgorithm}  where $\ell_{\max}$ is the number of iterations.
\begin{algorithm}[ht!]
\caption{BCD Alternating Algorithm to Solve \eqref{eq:opt1}}
\label{convexalgorithm}
\begin{algorithmic}[1]
	\Statex \textbf{Input:} $\ell_{\max}$, $P_{\mathrm{u},\max}$ $P_{\mathrm{d},\max}$, $P_{\mathrm{r},m\rr}$, $\gamma_{m\rr}$, $\forall\;m\rr$
	\Statex \textbf{Output: } $\mathbb{U}^\star$ $\mathbb{P}^\star$, $\mathbf{A}^\star$
	\State Set $\ell=0$; initialize $\mathbb{P}^{\paren{0}}$ and $\mathbf{A}^{\paren{0}}$
	\State Calculate $\mathbb{U}^{\paren{0}}$ with $\mathbb{P}^{\paren{0}}$ and $\mathbf{A}^{\paren{0}}$; find $\mathbb{W}^0$ with $\mathbb{U}^{\paren{0}}$ 
	\Repeat
	\For{$i=1,\cdots,\mathit{I}$}; 
	\State Calculate $\mathbf{P}^{\paren{\ell+1}}_{\mathrm{u},i}$ using the BB algorithm
	\EndFor
	\For{$m=1,\cdots,M$}
	\For{$j=1,\cdots,J$}
	\State Calculate $\mathbf{P}^{\paren{\ell+1}}_{\mathrm{d},mj}$ using the BB algorithm
	\EndFor
	\EndFor
	\For{$m\rr=1,\cdots,M\rr$} 
	\State 	Calculate $\mathbf{a}^{\prime}_{m\rr}$ using the BB algorithm
	\State Obtain $\mathbf{a}^{\paren{\ell+1}}_{m\rr}$ with Algorithm 3 in the second companion paper (Part II) \cite{liu2024codesigningpart2}
	\EndFor
	\State Update $\mathbb{U}^{\paren{\ell+1}}$ with $\mathbb{P}^{\paren{\ell+1}}$ and $\mathbf{A}^{\paren{\ell+1}}$; $\mathbb{W}^{\paren{\ell+1}}$ with $\mathbb{U}^{\paren{\ell+1}}$
	\Until $\ell>\ell_{\max}$
	\State \Return $\mathbb{U}^\star$, $\mathbb{P}^\star$, $\mathbf{A}^{\star}$
\end{algorithmic}\normalsize
\end{algorithm}

\subsection{Complexity and Convergence}
The computational cost for updating $\boldsymbol{\lambda}^{\paren{t}}_{\mathrm{u},i}$ with the BB method is $\mathcal{O}\paren{I}$ as opposed to $\mathcal{O}\paren{I^2}$ with Newton's method \cite{Boyd2020BB}. The low complexity of the BB method stems from the fact that each iteration incorporates the second-order derivative information without computing the Hessian approximates its inverse magnitude in contrast with Newton’s method \cite{BBSubgradient}.  In addition, the search direction of the BB method is the steepest descent, mirroring the Cauchy method but with a non-uniform step length, which renders the efficiency of the BB method \cite{burdakov2019stabilized}.  The cost to compute all of the elements in $\mathbb{P}$ with Algorithm \ref{convexalgorithm} is given as $\mathbb{O}\paren{I^2\paren{MM\cc}^3}$; see the second companion paper (Part II) \cite{liu2024codesigningpart2}. 

Literature has proved that the global convergence of the BB algorithm can be established for strictly convex quadratic functions, but there is only a guarantee of the local convergence for non-quadratics \cite{burdakov2019stabilized,BBSUmmwave}. As for Algorithm \ref{convexalgorithm}, the alternating sequence of iterates is not monotonically increasing hence it only reaches a local convergence, and different initialization points affect its local optimal values.
\section{D-ISAC Multi-Target Detection}
\label{sec:estimation}
This section presents how the D-ISAC system accomplishes multi-target detection and data association tasks. Despite the fact that target detection is a fundamental task for any radar system, there are limited works tackling multi-target localization with a distributed MIMO radar. We first apply a Neyman-Pearson (NP) hypothesis-based detector to retrieve each target's delay-Doppler information for each Tx-Rx channel in Section \ref{sec:detection}. One of the unique challenges faced by the distributed MIMO radar with $N_\target$ targets in the scene is the association ambiguity of the measurements obtained by each MIMO radar virtual antenna element. In Section \ref{sec:PDA}, we resort to the JPDA algorithm to associate measurements with their originating targets.
\subsection{Multi-Target Detection}
\label{sec:detection}
We assume that all the targets are well-separated at the $\ith{\paren{m\rr,n\rr}}$ radar channel, namely $n_{m\rr n_\target n\rr}\neq n_{m\rr n^\prime_\target n\rr}$ for $n_\target\neq n^\prime_\target$. The NP detector and the generalized likelihood ratio test (GLRT) detector are two of the most common detection strategies, where the former models the signal parameters as random variables with known probability density functions (PDFs), whereas the latter assumes the PDFs are unknown \cite{Cui2013Detector}. The NP detector obtains the optimal test statistic by maximizing the probability of detection ($P_{\mathrm{d}}$) with a certain probability of false alarm ($P_{\mathrm{fa}}$). In spite of the impracticability of the NP detector, we will use the
performance of the NP detector as an upper bound to compare the performance of beamforming and radar coding strategies in Section     To determine whether a range bin of the $\ith{\paren{m\rr,n\rr}}$ radar channel contains the $\ith{n_\target}$ target, we formulate a binary hypothesis test w.r.t. the cell under test (CUT), i.e.,
\begin{equation}
\label{eq: hypothesis}
\begin{cases}
\mathcal{H}_{\mathrm{0}}: & \mathbf{y}^{\mathrm{r}}_{m\rr n\rr} = \mathbf{y}^{\mathrm{r,in}}_{m\rr n\rr}\\
\mathcal{H}_{\mathrm{1}}: & \mathbf{y}^{\mathrm{r}}_{m\rr n\rr} = \mathbf{y}^{\mathrm{tr}}_{n_\target n\rr}+\mathbf{y}^{\mathrm{r,in}}_{n_\target n\rr}.
\end{cases}
\end{equation}
where $\mathcal{H}_{\mathrm{0}}$ corresponds to the absence of any targets and $\mathcal{H}_{\mathrm{1}}$ means the $\ith{n_\target}$ target is present. 
Define $\overline{\mathbf{y}}^{\mathrm{r}}_{n_\target n\rr} \triangleq \paren{\mathbf{R}^{\mathrm{r,in}}_{n_\target n\rr}}^{-\sfrac{1}{2}}\mathbf{y}^{\mathrm{r}}_{n_\target n\rr}$ and its CM $\mathbf{G}^{\mathrm{r}}_{n_\target n\rr}\triangleq\paren{\mathbf{R}^{\mathrm{r,in}}_{n_\target n\rr}}^{-\sfrac{1}{2}}\mathbf{R}^{\mathrm{r}}_{n_\target n\rr}\paren{\mathbf{R}^{\mathrm{r,in}}_{n_\target n\rr}}^{-\sfrac{1}{2}}$. We then rewrite \eqref{eq: hypothesis} as  
\begin{equation}
\begin{cases}
	\mathcal{H}_{\mathrm{0}}: & \overline{\mathbf{y}}^{\mathrm{r}}_{n_\target n\rr}\sim\mathcal{CN}\paren{\mathbf{0},\mathbf{I}}
	\\
	\mathcal{H}_{\mathrm{1}}: & \overline{\mathbf{y}}^{\mathrm{r}}_{n_\target n\rr}\sim\mathcal{CN}\paren{\mathbf{0},\mathbf{I}+\mathbf{G}^{\mathrm{r}}_{n_\target n\rr}},
\label{eq:NPHypoTest}
\end{cases}
\end{equation}\normalsize 
The eigendecomposition of $\mathbf{G}^{\mathrm{r}}_{n_\target n\rr}$ is $\mathbf{V}^{\mathrm{r}}_{n_\target n\rr}\mathbf{\Lambda}^{\mathrm{r}}_{n_\target n\rr}\mathbf{V}^{\mathrm{r}^\dagger}_{n_\target n\rr}$, where the columns of $\mathbf{V}^{\mathrm{r}}_{n_\target n\rr}\in\mathbb{C}^{K\times K}$ and the diagonal entries of $\mathbf{\Lambda}^{\mathrm{r}}_{n_\target n\rr}\triangleq\diag\bracket{\delta^{\mathrm{r},1}_{n_\target n\rr},\cdots,\delta^{\mathrm{r},m\rr}_{n_\target n\rr}}$ are, respectively, the eigenvectors and eigenvalues of $\mathbf{G}^{\mathrm{r}}_{n_\target n\rr}$ with $\delta^{\mathrm{r},m\rr}_{n_\target n\rr}$ the $\ith{k}$ eigenvalue. Manipulating $\overline{\mathbf{y}}^{\mathrm{r}}_{n_\target n\rr}$ with the Woodbury matrix identity attains the test statistic for \eqref{eq:NPHypoTest} as 
\begin{flalign}
T\paren{\overline{\mathbf{y}}^{\mathrm{r}}_{n_\target n\rr}}&=\sum_{n\rr=1}^{\mathit{N}\rr}T\paren{\overline{\mathbf{y}}^{\mathrm{r}}_{n_\target n\rr}}=\sum_{n\rr=1}^{\mathit{N}\rr}\overline{\mathbf{y}}^{\mathrm{r}^\dagger}_{n_\target n\rr}\paren{\mathbf{I}-\paren{\mathbf{G}^{\mathrm{r}}_{n_\target n\rr}+\mathbf{I}}^{-1}}\overline{\mathbf{y}}^{\mathrm{r}}_{n_\target n\rr}\nonumber\\
&=\sum_{n\rr=1}^{\mathit{N}\rr}\overline{\mathbf{y}}^{\mathrm{r}^\dagger}_{n_\target n\rr}\mathbf{V}^{\mathrm{r}}_{n_\target n\rr}\paren{\mathbf{\Lambda}^{\mathrm{r}^{-1}}_{n_\target n\rr}+\mathbf{I}}^{-1}\mathbf{V}^{\mathrm{r}^\dagger}_{n_\target n\rr}\overline{\mathbf{y}}^{\mathrm{r}}_{n_\target n\rr}
\end{flalign}\normalsize
Denote $\widehat{\mathbf{y}}^{\mathrm{r}}_{n_\target n\rr}=\mathbf{V}^{\mathrm{r}^\dagger}_{n_\target n\rr}\overline{\mathbf{y}}^{\mathrm{r}}_{n_\target n\rr}=\bracket{\widehat{y}^{\mathrm{r}}_{n_\target n\rr}\bracket{1},\cdots,\widehat{y}^{\mathrm{r}}_{n_\target n\rr}\bracket{\mathit{K}}}$. Then, the NP detector is \cite{Kay1993detection}
\begin{equation}
\label{eq: NPdetector}
T\paren{\overline{\mathbf{y}}^{\mathrm{r}}_{n_\target}}=\sum_{n\rr=1}^{\mathit{N}\rr}\sum_{k=1}^{\mathit{K}}\frac{\delta^{\mathrm{r},m\rr}_{n_\target n\rr}\lvert\widehat{y}^{\mathrm{r}}_{n_\target n\rr}\bracket{k}\rvert^2}{1+\delta^{\mathrm{r},m\rr}_{n_\target n\rr}}\underset{\mathrm{H}_0}{\overset{\mathrm{H}_1}{\gtrless}}\nu,
\end{equation}
where $\nu$ is the threshold selected to guarantee a certain  $P_{\mathrm{fa}}$. We apply \eqref{eq: NPdetector} to all range bins to retrieve the range information regarding the $N_\target$ targets. The Doppler information is then extracted using \eqref{eq:DopplerSpectrum} for each detected range bin.

\subsection{Data Association}
\label{sec:PDA}
JPDA incorporates all observations within a gated region about the predicted target state into the update of that target’s state. The contribution of each observation is determined by a probability-based weight. A given observation can also be used to update multiple targets’ states. In essence, JPDA averages over the data association hypotheses that have roughly comparable likelihoods and thus suffer from degradation in performance in a dense target environment. 

Upon iterating through each range-Doppler bin with the detection mechanism from Section \ref{sec:RadarSignalRadarRxs}, the $\ith{n\rr}$ radar Rx obtains $N_\target$ pairs of range-Doppler measurements. Each measurement is associated with at most one target, and all measurements are mutually independent.  associated with at most one target except for the clutter. All measurements are mutually independent, meaning that the number of measurements equals the number of targets. As mentioned in Section \ref{sec:RadarSignalRadarRxs}, we investigate measurement-to-target assignment using the PDA algorithm. We consider that the measurement is a simplified scenario where all the valid measurements collected at the $\ith{n\rr}$ Rx originate from the $N_\target$ targets after reflecting signals transmitted by $M\rr$ Txs.  After the matched filtering at each radar Rx, target-reflected signals originating from different radar and BS Txs are separated, and the Rx then associates each measurement with its corresponding Tx. The only ambiguity left is to assign each measurement with the correct target label across all radar Rxs. Therefore, the $\ith{q}$ delay-Doppler measurement extracted from the delay-Doppler plain corresponding to the $\ith{\paren{m\rr,n\rr}}$ Tx-Rx pair is $\mathbf{z}^q_{m\rr n\rr}$, where $q\in\braces{1,\cdots,N_\target}$. The \textit{unordered} set of measurements collected by the $\ith{m\rr,n\rr}$ radar channel is $\widetilde{\mathbb{Z}}_{m\rr n\rr}=\braces{\widetilde{\mathbf{z}}^1_{m\rr,n\rr},\cdots,\widetilde{\mathbf{z}}^{N_\target}_{m\rr,n\rr}}$ while the \textit{ordered} set of the $N_\target$ state vectors is $\mathbb{X}_{m\rr n_\target n\rr}=\braces{\mathbf{x}_{m\rr n_\target n\rr},\cdots,\mathbf{x}_{m\rr N_\target n\rr}}$. We assume that the measurements associated with target $n_\target$ are centered around its true delay and Doppler coordinates on the delay-Doppler plain and model the conditional distribution of  $\widetilde{\mathbf{z}}_{m\rr q n\rr}=\bracket{n^q_{m\rr,n\rr};\zeta^q_{m\rr,n\rr}}$ given $\mathbf{x}_{m\rr  n_\target n\rr}$ as
\begin{flalign}
&f\paren{\widetilde{\mathbf{z}}_{m\rr q n\rr}|\mathbf{x}_{m\rr  n_\target n\rr}}=\frac{\Delta_t\Delta_f}{2\pi\sigma^{\mathrm{t}}_{n_\target}\sigma^{\mathrm{p}}_{n_\target}}\nonumber\\
&\times e^{-\frac{1}{2}\bracket{\paren{\frac{n^q_{m\rr n\rr}-n_{m\rr n_\target n\rr}}{\sigma^{\mathrm{t}}_{n_\target}}}^2+\paren{\frac{p_{m\rr q n\rr}-f_{m\rr n_\target n\rr}}{\sigma^{\mathrm{p}}_{n_\target}}}^2}}
\label{eq:measurementmodel}
\end{flalign}
\normalsize
where $\Delta_t$ and $\Delta_f$ are the delay and Doppler shift resolutions. 

The composite model for measurement distribution is a mixture of each target component as follows 
\begin{equation}
f\paren{\widetilde{\mathbf{z}}_{m\rr q n\rr}|\mathbb{X}_{m\rr\target n\rr}}=\sum_{n_\target=1}^{N_\target}\phi_{m\rr n_\target n\rr}f\paren{\widetilde{\mathbf{z}}_{m\rr q n\rr}|\mathbf{x}_{m\rr  n_\target n\rr}},
\end{equation}
\normalsize
where $\phi_{m\rr n_\target n\rr}$ represents the weights of the $\ith{n_\target}$ target for all measurements observed by the $\ith{m\rr,n\rr}$ Tx-Rx pair, which is proportional to RCS $h^{\mathrm{r}}_{m\rr n_\target n\rr}$ and $\sum_{n_\target=1}^1\phi_{m\rr n_\target n\rr}=1$ for $\forall \braces{m\rr,n\rr}$. Then the probability that $\widetilde{\mathbf{z}}_{m\rr q n\rr}$ is generated by $\mathbf{x}_{m\rr  n_\target n\rr}$ is given as 
of each individual target component as follows \cite{Deming2009RangeDoppler}.
\begin{flalign}
&\mathrm{Pr}\paren{\mathbf{x}_{m\rr  n_\target n\rr}|\widetilde{\mathbf{z}}_{m\rr q n\rr}}=\frac{\phi_{m\rr n_\target n\rr}f\paren{\widetilde{\mathbf{z}}_{m\rr q n\rr}|\mathbf{x}_{m\rr  n_\target n\rr}}}{f\paren{\widetilde{\mathbf{z}}_{m\rr q n\rr}|\mathbb{X}}}\nonumber\\
&=\frac{\phi_{m\rr n_\target n\rr}\frac{e^{-\frac{1}{2}\bracket{\paren{\frac{n^q_{m\rr n\rr}-n_{m\rr n_\target n\rr}}{\sigma^{\mathrm{t}}_{n_\target}}}^2+\paren{\frac{p_{m\rr q n\rr}-p_{m\rr n_\target n\rr}}{\sigma^{\mathrm{p}}_{n_\target}}}^2}}}{\sum_{n^\prime_\target}e^{-\frac{1}{2}\bracket{\paren{\frac{n^q_{m\rr n\rr}-n_{m\rr n^\prime_\target n\rr}}{\sigma^{\mathrm{t}}_{n^\prime_\target}}}^2+\paren{\frac{p_{m\rr q n\rr}-p_{m\rr n^\prime_\target n\rr}}{\sigma^{\mathrm{p}}_{n^\prime_\target}}}^2}}}}{\sigma^{\mathrm{t}}_{n_\target}\sigma^{\mathrm{p}}_{n_\target}\sum_{n^\prime_\target}\frac{\phi_{m\rr n^\prime_\target n\rr}}{\sigma^{\mathrm{t}}_{n^\prime_\target}\sigma^{\mathrm{p}}_{n^\prime_\target}}},
\end{flalign}
\normalsize
which is the assignment likelihood between measurement $\widetilde{\mathbf{z}}_{m\rr q n\rr}$ and target $n_\target$. Then we construct the assignment likelihood matrix $\mathbf{L}_{m\rr,n\rr}$ for $N_\target$ targets and $N$ measurements, where the $\ith{\paren{n_\target,n}}$ element of $\mathbf{L}_{m\rr,n\rr}$ is $\mathrm{Pr}\paren{\mathbf{x}_{m\rr  n_\target n\rr}|\widetilde{\mathbf{z}}_{m\rr q n\rr}}$ \cite{Crouse2017Permanent}. Then we take advantage of the definition of matrix-permanent and calculate the association probability that the $\ith{n_\target}$ is assigned to the $\ith{q}$ measurement is 
\begin{equation}
\beta^{m\rr,n\rr}_{n,n_\target}=f\paren{\mathbf{x}_{m\rr  n_\target n\rr}|\widetilde{\mathbf{z}}_{m\rr q n\rr}}\frac{\mathrm{perm}\braces{\mathbf{L}^{n,n_\target}_{m\rr,n\rr}}}{\mathrm{perm}\braces{\mathbf{L}_{m\rr,n\rr}}},
\end{equation}
\normalsize
where $\mathbf{L}^{n,n_\target}_{m\rr,n\rr}$ is the matrix $\mathbf{L}_{m\rr,n\rr}$ removing row $n$ and column $n_\target$. We obtain the measurement vector $\mathbf{z}_{m\rr n_\target n\rr}=\widetilde{\mathbf{z}}^{n_{\max}}_{m\rr n\rr}$, where $n_{\max}=\arg\underset{n}{\max}\;\beta^{m\rr,n\rr}_{n,n_\target}$, for all Tx-Rx pairs.

To complete the localization and tracking with range and doppler measurements, we define the state space model of $\ith{n_\target}$ target in the $\ith{m}$ CPI (state $m$) as $\mathbf{x}^{m}_{\target,n_\target}=\bracket{x^{m}_{\target,n_\target},y^{m}_{\target,n_\target},\dot x^{m}_{\target,n_\target},\dot y^{m}_{\target,n_\target}}^\top$.
Assuming a nearly constant velocity discrete time kinematic target model\footnote{This model is a second-order model in that the discrete-time process noise is defined as a piecewise constant white sequence \cite{bar2004estimation}. } yields $\mathbf{x}^{m}_{\target,n_\target}=\mathbf{F}\mathbf{x}^{\paren{m-1}}_{\target,n_\target}+\mathbf{v}^{\paren{m-1}}_{\target,n_\target}$, where the state transition matrix is
\begin{equation}
\mathbf{F}=
\begin{bmatrix}
\mathbf{I}_{2\times2} & KT_{\mathrm{r}}\mathbf{I}_2\\
\mathbf{0}_{2\times2} & \mathbf{I}_{2\times2}
\end{bmatrix},
\end{equation}
and the process noise vector $\mathbf{v}^{\paren{m-1}}_{\target,n_\target}\sim\mathcal{N}\paren{\mathbf{0},\mathbf{Q}_{\target,n_\target}}$. The process noise covariance matrix is
\begin{equation}
\mathbf{Q}_{\target,n_\target}=
\begin{bmatrix}
\frac{\paren{KT\rr}^4}{4}\boldsymbol{\nu}_{\target,n_\target} & \frac{\paren{KT\rr}^3}{2}\boldsymbol{\nu}_{\target,n_\target}\\
\frac{\paren{KT\rr}^3}{2}\boldsymbol{\nu}_{\target,n_\target} & \paren{KT\rr}^2\boldsymbol{\nu}_{\target,n_\target}
\end{bmatrix},
\end{equation}
where $\boldsymbol{\nu}_{\target,n_\target}=\diag\paren{\bracket{\nu^2_{x,n_\target},\nu^2_{y,n_\target}}}$ with $\nu^2_{x,n_\target}$ ($\nu^2_{y,n_\target}$) the process noise intensity on the x-axis (y-axis). The measurements are taken at each receiver and overall, measurement vector at a time instant $m$
of radar Rx $n\rr$ from target $n_\target$ are given by
\begin{equation}
\mathbf{z}^m_{m\rr n_\target n\rr}=\mathbf{b}_{m\rr n\rr}\paren{\mathbf{x}^{m}_{\target,n_\target}}+\mathbf{e}_{m\rr n\rr}
\end{equation}\normalsize
where $\mathbf{e}_{m\rr n\rr}\sim\mathcal{N}\paren{\mathbf{0},\boldsymbol{\Omega}_{m\rr n\rr}}$ is the time-invariant measurement error vector which is independent across different receivers of the $\ith{\paren{m\rr n\rr}}$ radar path and $\boldsymbol{\Omega}_{m\rr n\rr}$ is the measurement noise covariance matrix; $\mathbf{b}_{m\rr  n\rr}\paren{\cdot}=\bracket{b_{\tau,m\rr n\rr}\paren{\cdot},b_{f,m\rr n\rr}\paren{\cdot}}^\top$ 

Since radar measurement errors are independent across different receivers, the covariance matrix of $\mathbf{e}_{m\rr n\rr}$ is given by
\begin{equation}
R^m = \mathbb{E}[\mathbf{e}_{m\rr n\rr}\mathbf{e}_{m\rr n\rr}^\top ]= \text{diag}(R_{m\rr n\rr}^m, \ldots, R_{m\rr n\rr}^m)
\end{equation}

where the covariance matrix \(R_{m\rr n\rr}\) is computed from the corresponding CRLB matrix \(C_{m_rn_r}^m\) as
\begin{equation}
R_{m_rn_r}^m = \Gamma C_{m_rn_r}^m \Gamma^\top
\end{equation}
where $\Gamma$ is the diagonal element of transformation from delay, doppler to range and range-rate, respectively.

From the measurement noise equation, $\mathbf{b}_{m\rr n\rr}$ is the nonlinear observation vectorial function. Due to the non-linearity of $\mathbf{b}_{m\rr n\rr}$, a nonlinear tracking algorithm must be used. Therefore, we perform the target state update using the extended Kalman filter (EKF). The error covariance matrix depends on the $R_{m_rn_r}^m$ depending on the SNR $\boldsymbol\eta_{m_rn_r}$. The probability of target detection at $\ith{n_\target}$ receiver at time instant $m$ is

\begin{equation}
P_{d_{m_rn_r}}^m = P_{f_{m_rn_r}}^{(1/1+ \boldsymbol\eta_{m_rn_r})}
\end{equation}

The SNR $\boldsymbol\eta_{m_rn_r}$ varies with target motion and is directly related to the bistatic geometry of the transmitter-receiver location. 
Measurements that are accurate and reflect the actual target (when it is identified) are not the only data collected by each sensor. Unwanted noise is also detected due to background disturbances. These false readings are considered to have a uniform spatial distribution within the measurement domain and are temporally independent. The frequency of these false positives is represented by a Poisson distribution. The function describing the likelihood of observing a certain number of false positives within a given volume \( V \) is outlined below:

\begin{equation}
Po(g) = e^{-\lambda V} \frac{(\lambda V)^g}{g!}
\end{equation}

The prediction steps of the EKF w.r.t. the $\ith{n_\target}$ target given by,
\begin{flalign}
\mathbf{x}^{m|m-1}_{\target,n_\target}=&\mathbf{F}\mathbf{x}^{m-1}_{\target,n_\target}\\
\mathbf{P}^{m|m-1}_{\target,n_\target}=&\mathbf{F}\mathbf{P}^{m-1}_{\target,n_\target}\mathbf{F}^\top+\mathbf{Q}_{\target,n_\target}.
\end{flalign}
\normalsize
The EKF innovation given the measurement yields the current localization of the $\ith{n_\target}$ target 
\begin{flalign}
\mathbf{x}^{m}_{\target,n_\target}=\mathbf{x}^{m|m-1}_{\target,n_\target}+\mathbf{K}^{m}_{\target,n_\target}\overline{\mathbf{z}}^m_{m\rr n_\target n\rr}
\end{flalign}
\normalsize
where $\overline{\mathbf{z}}^m_{m\rr n_\target n\rr}=\mathbf{z}^m_{m\rr n_\target n\rr}-\mathbf{b}_{m\rr n\rr}\paren{\mathbf{x}^{m|m-1}_{\target,n_\target}}$ is the filter residual of $\ith{m}$ state, The  Kalman gain is given by,
\begin{equation}
\mathbf{K}^{m}_{\target,n_\target}=\mathbf{P}^{m|m-1}_{\target,n_\target}\paren{\mathbf{B}^{m}_{m\rr n_\target n\rr}}^\top\paren{\mathbf{S}^{m}_{\target,n_\target}}^{-1}
\end{equation}

Here, $\mathbf{B}^{m}_{m\rr n_\target n\rr}$ is the Jacobian of $\mathbf{b}_{m\rr n\rr}$ evaluated at $\mathbf{x}^{m|m-1}_{\target,n_\target}$, and the residual covariance matrix is 
\begin{equation}
\mathbf{S}^{m}_{\target,n_\target}=\mathbf{B}^{m}_{m\rr n_\target n\rr}\mathbf{P}^{m|m-1}_{\target,n_\target}\mathbf{b}^{m^\top}_{m\rr n_\target n\rr}+\boldsymbol{\Omega}_{m\rr n\rr}
\end{equation}

In the multi-target scenario, the measurements from $\ith{n_\target}$ target during each iteration are not independent. So joint associations between the transmitters and targets need to be addressed for correct receiver measurement. JPDA enumerates the measurement to target association probabilistically, and the target states are estimated by their marginal association probability. Moreover, the JPDA filter will resolve the ambiguity among measurements, targets, and transmitters for a specific receiver. Therefore, this association is a three-dimensional association that has a higher computation cost. Therefore, we followed another modified approach where we consider the super-target formation mentioned in \cite{7527957} \cite{10.1117/12.829599} \cite{6809941}, which turns the 3D association into a 2D association problem. We consider supertarget $\tilde{\tau} = \{n_t, m_r\}$, which is a hypothetical target consisting of a pair of target $n_t$ and transmitter $m_r$ for any specific receiver. As the number of transmitters and targets grows, the association between the measurement and the target increases accordingly. Gate grouping is required for multiple targets \cite{10.1117/12.848895}. The valid measurement is denoted by the set of gated measurements at time $m$ concerning supertarget $\tilde{\tau}_{m_rn_r}$, that is, the ${m_rn_tn_r}\ith{}$ measurement of $\mathbf{z}^m_{\tilde{\tau}n_r}$:

\begin{equation}
\mathbf{z}^m_{\tilde{\tau},n_r}=\mathbf{z}^m_{m\rr n_\target n\rr}\in \mathbf{Z}^m:(\mathbf{z}^m_{m\rr n_\target n\rr}-\mathbf{b}_{m\rr n\rr}\paren{\mathbf{x}^{m|m-1}_{\target,n_\target}, \mathbf{Tx}_{m_r}})^{\top}\\
\end{equation}
\begin{equation}
 (\mathbf{S}^m_{\target,n_\target{m_r}})^{-1}(\mathbf{z}^m_{m\rr n_\target n\rr}-\mathbf{b}_{m\rr n\rr}\paren{\mathbf{x}^{m|m-1}_{\target,n_\target}, \mathbf{Tx}_{m_r}}) < \mathbb{Th\}}
\end{equation}
with the predicted measurement $\mathbf{b}_{m\rr n\rr}\paren{\mathbf{x}^{m|m-1}_{\target,n_\target}, \mathbf{Tx}_{m_r}})$ and its associated covariance $\mathbf{S}^m_{\target,n_\target{m_r}}$ with respect to supertarget $\tilde{\tau} = \{n_t, m_r\}$, and $\mathbb{Th}$ is the gating threshold.
Next, we consider a track to be lost if, over several consecutive scans, no measurements are found within the designated target gates or if the gate size becomes excessively large \cite{10.1117/12.829599}.

\begin{equation}
p(\chi^m_{n_t}, \mathbf{x}^m_{n_t} | \mathbf{Z}^{m-1}) = P(\chi^m_{n_t} | \mathbf{Z}^{m-1})p(\mathbf{x}^m_{n_t} | \chi^m_{n_t}, \mathbf{Z}^{m-1})
\end{equation}
\begin{equation}
p(\mathbf{x}^m_{n_t} | \chi^m_{n_t}, \mathbf{Z}^{m-1}) = \mathcal{N}(\mathbf{x}^m_{n_t}; \hat{\mathbf{x}}^{m|m-1}_{n_t}, \mathbf{P}^{m|m-1}_{n_t}).
\end{equation}
Let the term $\xi_j$ represent the $j^{th}$ fusion junction event (FJE), with $T_0(\xi_j)$ indicating the collection of supertargets that are not linked to any measurement, and $T_1(\xi_j)$ representing the collection of supertargets that are linked to exactly one measurement in the context of the FJE $\xi_j$. The posterior probability for the FJE $\xi_j$ is calculated as follows:

\begin{align}
P(\xi_{j} | Z^m) = c^m\prod_{\tilde{\tau} \in T_{0}(\xi_{j})}(1 - P_{D}P_{G}P(\chi^m_{\tau} | Z^{m-1})) \times \prod_{\tilde{\tau} \in T_{1}(\xi_{j})}(P_{D}P_{G}P(\chi^m_{\tau} | \mathbf{Z}^{m-1})\frac{p^m_{{\tilde{\tau},n_tm_r}}}{\rho^m_{n_tm_r}}),
\end{align}
with $P_G$ as the gating probability and $c^m$ as the normalization constant. The measurement likelihood allocated to a supertarget in a fusion junction event is given by
\begin{equation}
\hat{p}_{n_t,j}^{m} = \frac{N(\mathbf{Z}_{n_t,j}^{m}; \mathbf{b}_{m\rr n\rr}^{m}(\hat{\mathbf{x}}_{n_t}^{m|m-1}, (\mathbf{S}^m_{\target,n_\target{m_r}}))}{P_G}.
\end{equation}

The set of fusion junction events allocating a measurement to a supertarget is denoted by $\theta(\hat{\tau}, n_tm_r)$, identifying if a measurement $i$ detects a supertarget $\hat{\tau}$ at time $m$. The probability that no measurement detects the supertarget is
\begin{equation}
P({\chi_{\tilde{\tau}}}^{m,0} | \mathbf{Z}^m) = \sum_{\xi_j \in \theta(\tilde{\tau},0)} P(\xi_j | \mathbf{Z}^m),
\end{equation}
and the probability that a measurement detects the supertarget and confirms its existence is

\begin{equation}
P({\chi_{\tilde{\tau}}}^{m}, {\chi_{{\tilde\tau} n_tm_r}}^{m}, | \mathbf{Z}^m) = \sum_{\theta_j \in \theta(\tilde{\tau},n_tm_r)} P(\theta_j | \mathbf{Z}^m),
\end{equation}
resulting in the supertarget existence probability as
\begin{equation}
P({\chi_{\tilde{\tau}}}^{m} | \mathbf{Z}^m) = \sum_{n_tm_r \geq 0} P({\chi_{\tilde{\tau}}}^{m}, {\chi_{{\tilde\tau} n_tm_r}}^{m} | \mathbf{Z}^m).
\end{equation}

The data association probability for a supertarget is
\begin{equation}
\beta_{{\tilde\tau},n_tm_r} = P({\chi_{{\tilde\tau} n_tm_r}}^{m}  | {\chi_{\tilde{\tau}}}^{m} , \mathbf{Z}^m) = \frac{P({\chi_{{\tilde\tau} n_tm_r}}^{m}, {\chi_{\tilde{\tau}}}^{m} | \mathbf{Z}^m)}{P({\chi_{\tilde{\tau}}}^{m}  | \mathbf{Z}^m)}, \quad n_tm_r \geq 0.
\end{equation}

For track state updates, combining each track and transmitter forms a supertarget. The track state for each $n_t$ at time $m$ integrates all originating supertargets, with the probability density function modeled as a Gaussian distribution, encapsulating the mean and covariance updated by measurements related to the supertarget. The extended Kalman filter update is applied for state refinement based on predicted measurements.

The probability density function of the track $n_t$ trajectory state is assumed to be a single Gaussian distribution.

\begin{equation} p(\chi_{n_t}^{m},\mathbf{x}_{n_t}^{m}\vert \mathbf{Z}^{m})=P(\chi_{n_t}^{m}\vert \mathbf{Z}^{m})\mathbf{p}(\mathbf{x}_{n_t}^{m}\vert \mathbf{Z}^{m}).  \end{equation}

\begin{equation}
P(\mathbf{x}_{n_t}^{m}|\mathbf{Z}^m) = \mathcal{N}(\mathbf{x}_{n_t}^{m}; {\hat{\mathbf{x}}_{n_t}}^{m|m}, {\mathbf{p}_{n_t}}^{m|m}),
\end{equation}
where
\begin{equation} 
{\hat{\mathbf{x}}_{n_t}}^{m \vert m}= \sum_{\tilde{\tau}\in \text{E}(n_t)} \sum_{j=0}^{N_{n_t}^{m}}\tilde{c}^{m}\beta_{{\tilde\tau},j}\hat{\mathbf{x}}^{m\vert m}_{{\tilde\tau},j}  
\end{equation}
\begin{equation}
{\mathbf{p}_{n_t}^{m|m}} = \sum_{\tilde\tau \in E(\tau)} \sum_{j=0}^{N_{\tilde\tau}^m} \tilde{c}^{m} \beta_{{\tilde\tau},j} \left(\mathbf{p}_{\tilde\tau,j}^{m|m} + 
\hat{\mathbf{x}}^{m\vert m}_{{\tilde\tau},j}
(\hat{\mathbf{x}}^{m\vert m}_{{\tilde\tau},j})^\top \right) - {\hat{\mathbf{x}}_{n_t}}^{m \vert m}({\hat{\mathbf{x}}_{n_t}}^{m \vert m})^\top.
\end{equation}

where $\tilde{\mathcal{C}}^m$ is the renormalized factor which satisfies
\begin{equation}
\sum_{r \in E(n_t)} \sum_{j=0}^{N_{n_t}^m} \tilde{\mathcal{C}}^m \beta_{{\tilde\tau},j} = 1.
\end{equation}
The mean and covariance updated by measurement $\mathbf{z}^m_{\tilde{\tau},n_r}$ with respect to supertarget $\tilde\tau$ are calculated by
\begin{equation}
\left[\hat{\mathbf{x}}^{m\vert m}_{{\tilde\tau},n_tm_r}, \mathbf{p}_{\tilde\tau,n_rm_r}^{m|m}\right] = 
\begin{cases}
\left[({\hat{\mathbf{x}}_{n_t}}^{m \vert m-1})^\top \mathbf{p}_{\tilde\tau}^{m|m-1}\right], & j = 0 \\
\text{EKFu}\left({\hat{\mathbf{x}}}_{n_t}^{m|m-1}, R^{m}, \mathbf{Z}_{\tilde\tau}^{m|m-1}, \mathbf{b}_{m\rr n\rr}^{m}\right), & j > 0,
\end{cases}
\end{equation}
where $\text{EKFu}$ is the extended Kalman filter update procedure, and the predicted measurement function is $\mathbf{b}_{m\rr n\rr}^{m} = \mathbf{b}_{m\rr n\rr}^{m}(\hat{\mathbf{x}}_{\tilde\tau}^{m|m-1}, \mathbf{Tx}_{m_r}).$
The calculated posterior probability of a target's presence, related to track $n_t$ at the instance $m$ is determined by

\begin{equation}
P(\hat{\mathbf{\chi}}_{\tilde\tau}^{m} | \mathbf{Z}^m) = \sum_{\tilde\tau \in E(n_t)} P(\hat{\mathbf{\chi}}_{\tilde\tau}^{m} | \mathbf{Z}^m)/N_{n_t}.
\end{equation}

\section{Numerical Experiments}
\label{sec:numerical}
We evaluate the proposed algorithm's performance, target detection, and localization performance with the considered D-ISAC system. Throughout this section, we assume the following parameters, unless otherwise stated: $M\rr=N\rr=M=4$, $I=J=2$, $K=16$, and $L=32$; the CSCG noise variances are $\sigma^2_{\mathrm{r}}=\sigma^2_{\mathrm{u}}=\sigma^2_{\mathrm{d}}=0.01$; $\mathit{P}_\textrm{U}=1$ and DL $\mathit{P}_\textrm{d}=2$; radar Tx power $\mathit{P}_{\textrm{r},m\rr}=1$ and PAR levels $\gamma_{m\rr}=2$ for all $\braces{m\rr}$; define the signal-to-noise ratios (SNRs) associated with the MIMO radar, $\ith{j}$ DL UE, and $\ith{i}$ UL UE as $\mathrm{SNR}_{\textrm{r}}=\sfrac{P_{\textrm{r},m\rr}}{\sigma^2_{\textrm{r}}}$, $\mathrm{SNR}_{\textrm{DL}}=\sfrac{P_{\mathrm{d}}}{\sigma^2_{\textrm{d}}}$, and $\mathrm{SNR}_{\textrm{UL}}=\sfrac{P_{\textrm{u},i}}{\sigma^2_{\mathrm{u}}}$ \cite{Luo2011IterativeWMMSE}.  the elements of $g_{m\rr n_\target n\rr}$ $\mathbf{h}_{\textrm{u},i}$, $\mathbf{g}_{\textrm{},i}$ $\mathbf{H}_{i,\textrm{B}}$, $\mathbf{H}_{i,j}$, $\boldsymbol{\alpha}_{\textrm{Bm},n\rr}$, and $\boldsymbol{\alpha}_{i,n\rr}$ are drawn from $\mathcal{CN}\paren{0,1}$
We model the self-interfering channel $\HBB$ as $\mathcal{CN}\paren{\sqrt{\frac{\sigma^2_{\mathrm{FD}}K_{\B}}{1+K_{\B}}}\widehat{\mathbf{H}}^{\mathrm{SR}}_{\mathrm{u}},\frac{\sigma^2_{\mathrm{FD}}}{1+K_{\B}}\mathbf{I}_{MM\cc}\otimes\mathbf{I}_{MM\cc}}$, where $\sigma^2_{\mathrm{FD}}$ is the SI attenuation coefficient that characterizes the effectiveness of SI cancellation \cite{Schober16distributedFD}, the Rician factor $K_{\B}=1$, and $\widehat{\mathbf{H}}^{\mathrm{SR}}_{\mathrm{u}}\in\mathbb{C}^{MM\cc\times MM\cc}$ is an all-one matrix \cite{FD_WMMSE}. The numbers of iterations for the BB and BCD-Alternating optimization algorithms are $t_{\max}=100$ and $\ell_{\textrm{max}}=100$. We use uniform weights  $\alpha^\textrm{u}_{i}= \alpha^{\textrm{d}}_{j}=\alpha^\textrm{r}_{m\rr n_\target n\rr}$ for all $\braces{m\rr, n_\target, n\rr,i,j}$. Throughout this section, we initialize Algorithm \ref{convexalgorithm}
 with the efficient initialization approach for $\mathbf{v}_{\mathrm{d},j}$ detailed in the second companion paper (Part II) \cite{liu2024codesigningpart2}.
\subsection{Convergence analysis}
\begin{figure}[t]
	\centering
	\includegraphics[width=1.0\columnwidth]{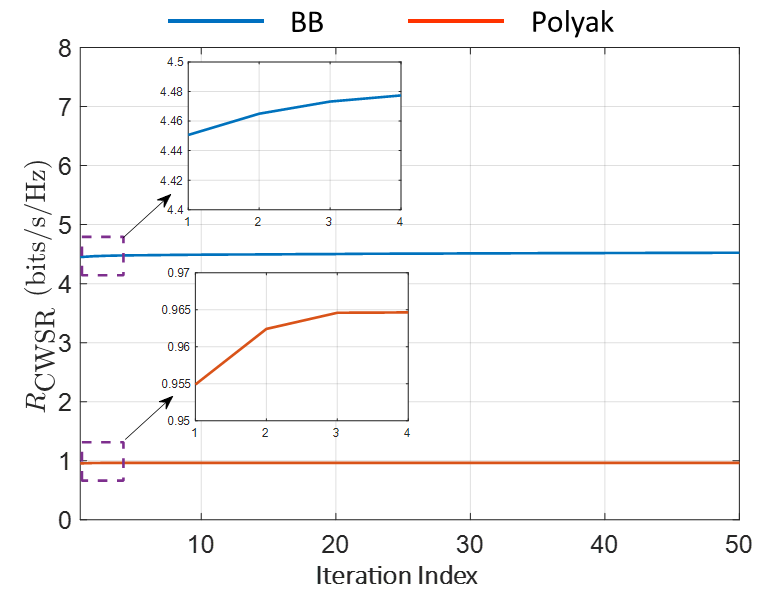}
	\caption{Convergence curves of Algorithm \ref{convexalgorithm} given the step-size rule determined by the BB algorithm and Polyak's rule.
	}
	\label{fig:convergence}
\end{figure}
\figurename{~\ref{fig:convergence}} presents the convergence behavior of Algorithm \ref{convexalgorithm} with two step-size rules: the BB algorithm and the Polyak's rule, which shows that Algorithm \ref{convexalgorithm} achieves a rapid convergence using both step-size rules because the co-phasing technique adopted for $\mathbf{v}_{\mathrm{d},j}$ uses the channel phase information, which aligns with initialization approach. It is also noted that the BB algorithm yields improved performance over that of Polyak's rule.
\begin{figure}[t]
	\centering
	\includegraphics[width=1.0\columnwidth]{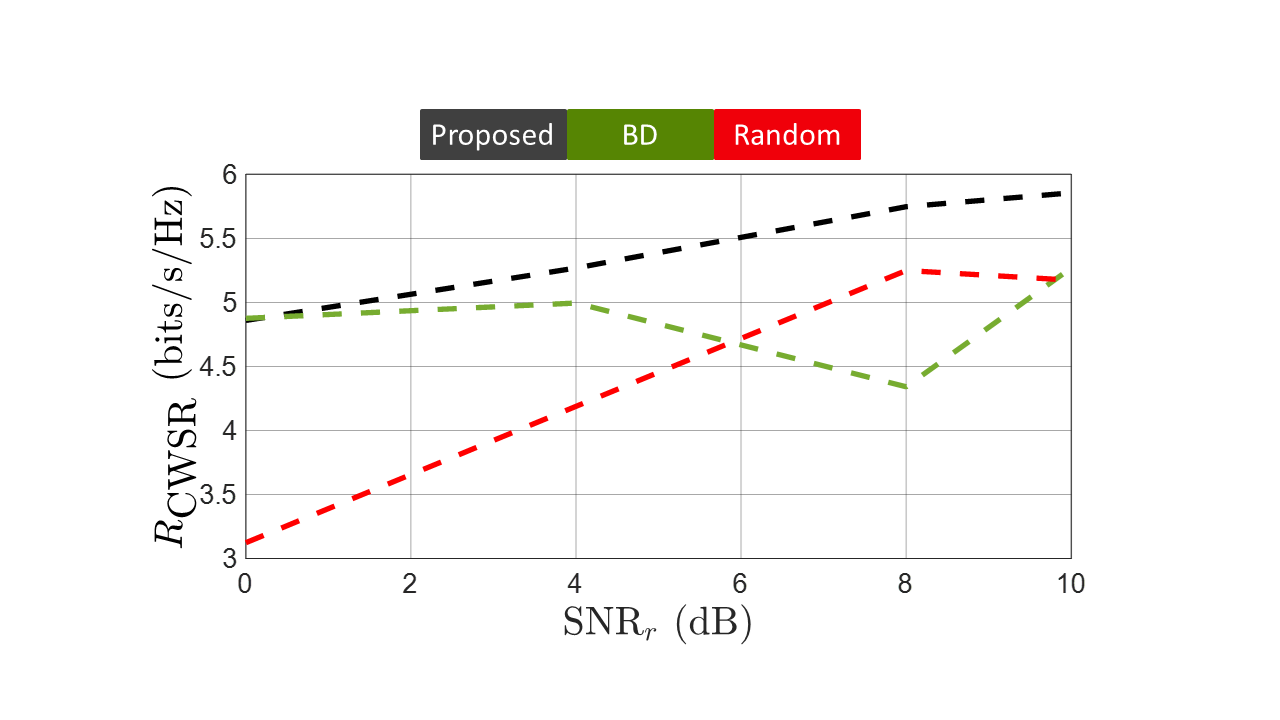}
	\caption{Proposed D-ISAC design approach compared with the conventional communications precoding and radar coding techniques given CSI errors.}
	\label{fig:system_vs_SNR}
\end{figure}
\subsection{D-ISAC overall system evaluation}
\figurename{\ref{fig:system_vs_SNR}}  demonstrates the overall D-ISAC system performance measured by \eqref{eq: WSR} and the robustness of Algorithm \ref{convexalgorithm} with channel estimation errors. We model the estimated channel vectors as $\widehat{\mathbf{h}}={\mathbf{h}}+\Delta$, where $\Delta\sim\mathcal{CN}\paren{\mathbf{0},\eta^2_{\mathrm{CSI}}\mathbf{I}}$ and $\mathbf{h}$ is referred to any of the small fading channel vectors introduced in Section \ref{sec: system_channel} and $\eta^2_{\mathrm{CSI}}=0.1$ for this example. We also compare the proposed BCD-Alternating algorithm with conventional strategies, where we apply the block-diagonal (BD) beamforming technique to the DL beamformer $\mathbf{v}_{\mathrm{d},j}$ and the random radar coding scheme to the radar code matrix $\mathbf{A}$ (see the first companion paper [Part I] \cite{liu2024codesigningpart1}), where the BD approach only applies to the DL beamforming; the proposed DL and UL beamforming is used with the random radar coding. The proposed D-ISAC design approach displays its overall robustness given the channel uncertainty and system-level advantage over other compared design approaches.
\subsection{FD communications}
\begin{figure}[t]
	\centering
	\includegraphics[width=1.0\columnwidth]{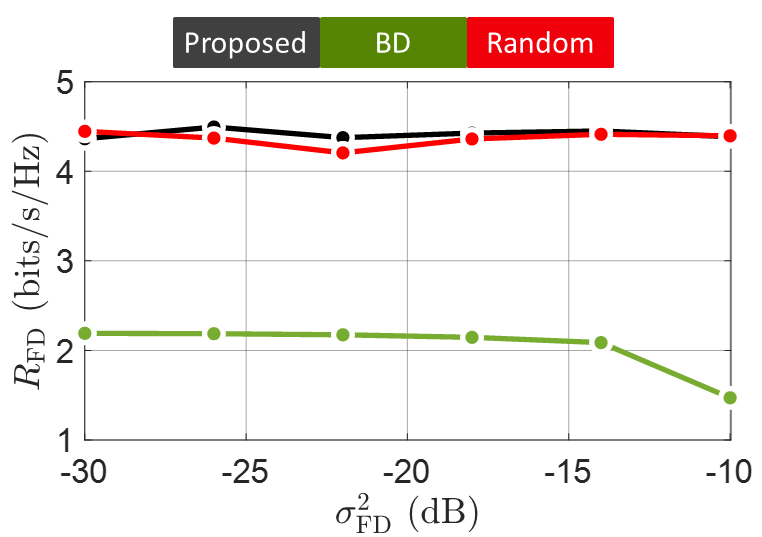}
	\caption{Proposed D-ISAC design approach compared with the conventional communications precoding and radar coding techniques given CSI errors.}
	\label{fig:system_vs_FD}
\end{figure}
We then evaluate the impact of the FD communications on the FD C-RAN system, where \figurename{~\ref{fig:system_vs_FD}} depicts $R_{\mathrm{FD}}$ against the SI attenuation level $\sigma^2_{\mathrm{FD}}$ ranging from $-30$ to $-10$ dB, where $R_{\mathrm{FD}}$ represents the weighted achievable rate of the FD-CRAN system given by the second term on the right-hand-side of \eqref{eq: WSR}. Similar to the previous example, we consider the BD beamforming and radar random coding as benchmarks for our algorithm. As expected, the stronger the SI cancellation, the higher the FD C-RAN system achievable rate becomes. The FD C-RAN systems using the UL and DL beamformers based on our proposed algorithm outperform the one using the BD method for DL beamforming. Since the communications beamforming is applied to the radar random coding, we observe that the black and red curves share a similar trend in 

\subsection{Radar target data association}
To evaluate the data association approach described in Section \ref{sec:PDA}, we consider three sets of MIMO radar antenna array configurations. Further, we simulate the association probability for numerous targets in a 2D Cartesian plain.

We generate measurements following \eqref{eq:measurementmodel} for all Tx-Rx channels, where $P_{\mathrm{d}} = 0.9$ and $P_{\mathrm{f}} = 0.001$.  
\begin{figure}[t]
	\centering
	\includegraphics[width=1\columnwidth]{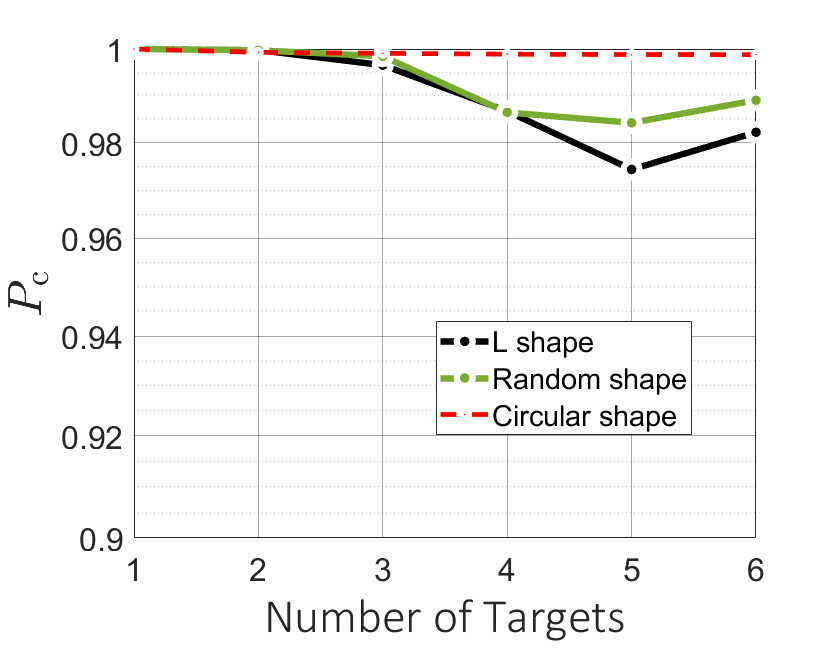}
	
	\caption{Probability of correct association for different antenna geometries and number of targets.}
	\label{fig:Pc}
\end{figure}

\begin{figure}[t]
	\centering
	\includegraphics[width=1\columnwidth]{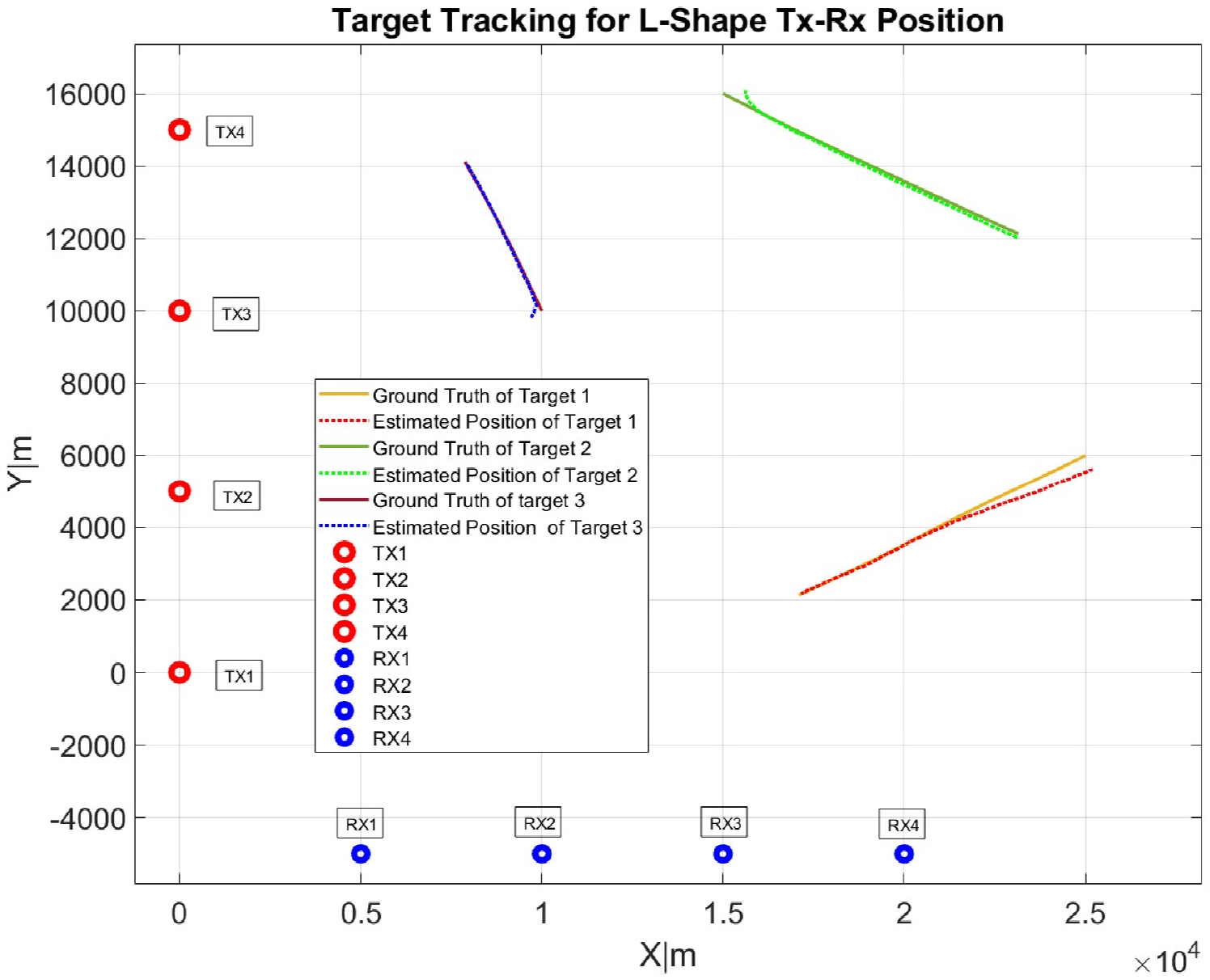}
	
	\caption{Ground truth and estimated location of three targets for L-shape Tx-Rx configuration.}
	\label{fig:Lshape}
\end{figure}
To quantify the performance of the proposed data association scheme, the probability of correct data association is defined as $P_\mathrm{c}=d_{\mathrm{c}}/D$, where $d_{\mathrm{c}}$ is the number of measurements associated to the correct targets, and $D$ is the number of total valid measurements. We compute $P_\mathrm{c}$ by varying the number of targets for three different popular array configurations (Circular, L-shape and Random) in Figure~\ref{fig:Pc}, where each curve is averaged over $2000$ MC realizations, in each realization $\paren{x_{\target,n_\target},y_{\target,n_\target}}$ is sampled randomly within a circle of radius $300$ $Km$ and $\sqrt{\dot x^2_{\target,n_\target}+\dot y^2_{\target,n_\target}}$ is uniformly generated within $\bracket{0,50}$ for all $n_\target$. In Figure~\ref{fig:Pc}, it can be noticed that among the three configurations, the circular configuration is the best in terms of the probability of correct association due to its $360^\circ$ coverage, and as expected, its performance degrades with the increase in the number of targets.

\begin{figure}[t]
	\centering
	\includegraphics[width=1\columnwidth]{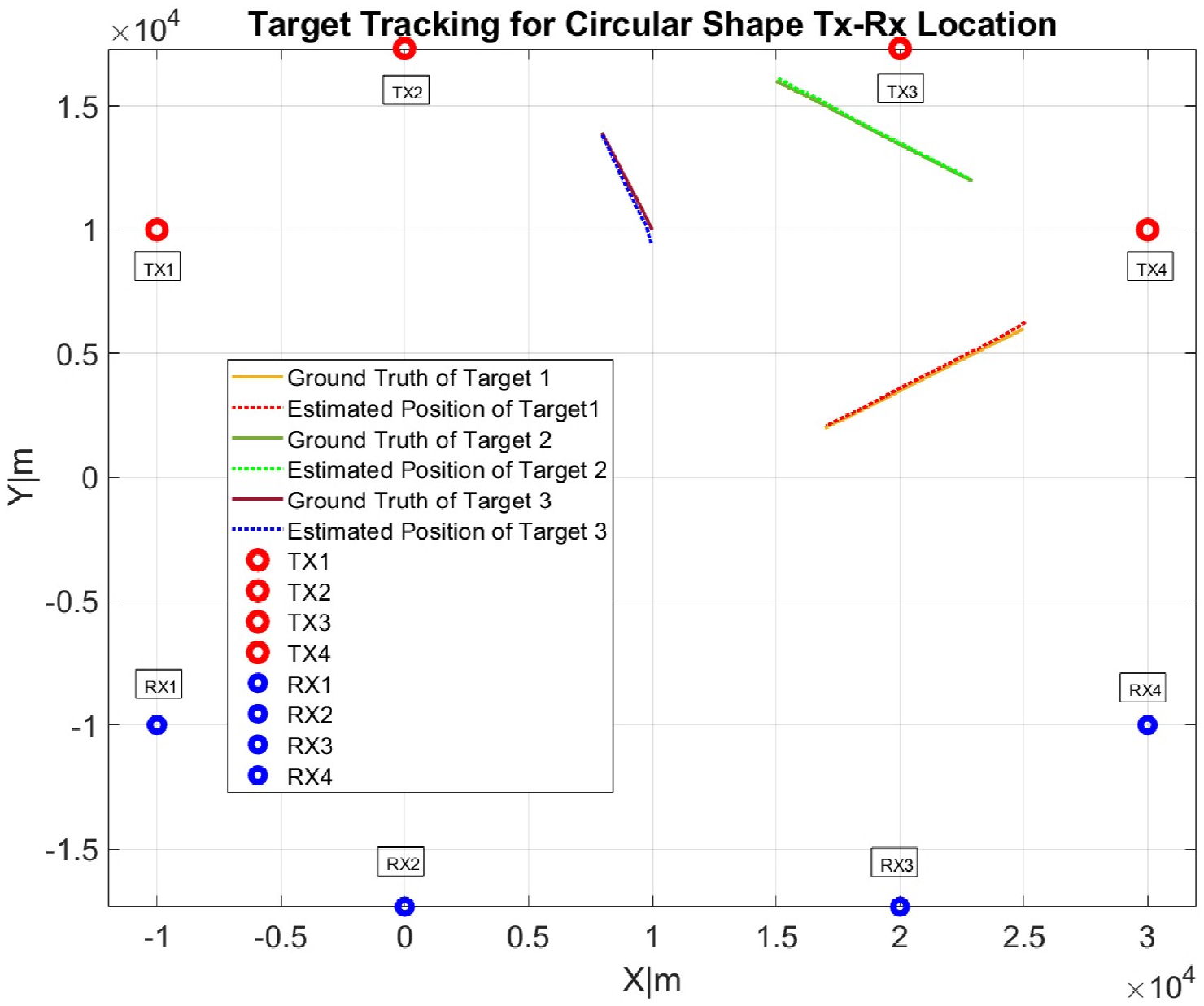}
	
	\caption{Ground truth and estimated location of three targets in Circular Tx-Rx configuration.}
	\label{fig:circle}
\end{figure}
\begin{figure}[t]
	\centering
	\includegraphics[width=1.0\columnwidth]{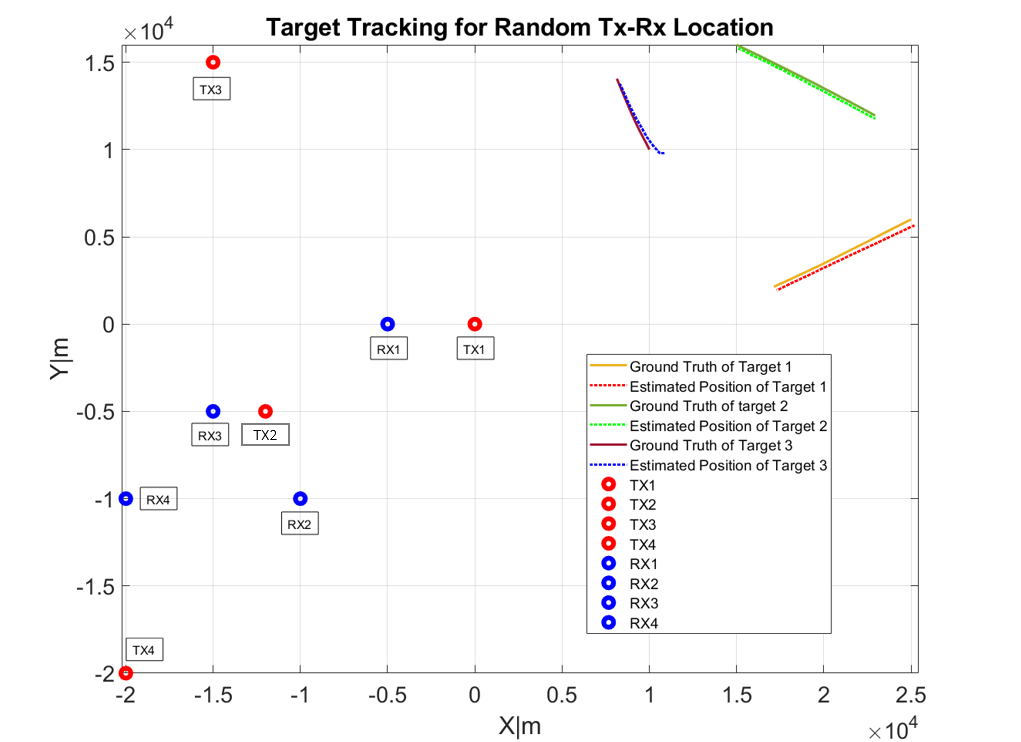}
	\caption{Ground truth and estimated location of three targets for Random Tx-Rx configuration. }
	\label{fig: random_pdf.pdf}
\end{figure}

The performance efficacy of our distributed radar system for multi-target tracking was rigorously assessed via a series of numerical simulations. The experimental setup encompassed the tracking of three distinct targets, utilizing an array of four transmitters and four receivers. These components were strategically positioned in a variety of geometric configurations within a two-dimensional Cartesian coordinate system. Figures~\ref{fig:Lshape}, \ref{fig:circle} and ~\ref{fig: random_pdf.pdf} depict the simulated tracking scenario where Txs and Rxs are distributed in an L-shape, circular, and random configuration, respectively. The radar transmission frequency was set to 12 GHz for all the Txs with a 200 ms PRI. The $P_{\mathrm{fa}}$ was set to 0.001. The term \( T \) denotes the interval of sampling, which we set to 200 ms. The probability of detection is assumed to be equal across every Tx-Rx pair. 

For the L-shape, the Txs and Rxs are located as follows: $$\mbox{Tx}_m = [0\quad (m-1)5]^\top; \quad m = 1,2,3,4 \quad \mbox{and} \quad \mbox{Rx}_n = [5n \quad -5]^\top; \quad n = 1,2,3,4\,.$$

For the circular configuration, the transmitters and receivers are located as follows:
$$\mbox{Tx}_1 = [-10\quad 10]^\top; \quad \mbox{Tx}_2 = [0\quad 17.32]^\top; \quad \mbox{and} \quad \mbox{Tx}_3 = [20 \quad 17.32]^\top; \quad \mbox{Tx}_4 = [30 \quad 10]^\top.$$

$$\mbox{Rx}_1 = [-10\quad -10]^\top; \quad \mbox{Rx}_2 = [0\quad -17.32]^\top; \quad \mbox{and} \quad \mbox{Rx}_3 = [20 \quad -17.32]^\top; \quad \mbox{Rx}_4 = [30\quad -10]^\top.$$

For the random configuration, the transmitters and receivers are located as follows:
$$\mbox{Tx}_1 = [0\quad 0]^\top; \quad \mbox{Tx}_2 = [-10\quad -5]^\top; \quad \mbox{and} \quad \mbox{Tx}_3 = [-15 \quad -5]^\top; \quad \mbox{Tx}_4 = [-20 \quad -20]^\top.$$

$$\mbox{Rx}_1 = [-5\quad 0]^\top; \quad \mbox{Rx}_2 = [-10\quad -10]^\top; \quad \mbox{and} \quad \mbox{Rx}_3 = [-15 \quad -5]^\top; \quad \mbox{Rx}_4 = [-20\quad -10]^\top.$$
The targets follow a nearly constant velocity motion model, with initial states for target 1, target 2, and target 3 as, respectively, $$[25\quad 6\, -0.4\, -0.2]^\top; \quad [15\quad  16 \quad 0.4\quad -0.2]^\top; \quad [10\quad 10\quad -0.1\quad 0.2]^\top\,.$$ 
We consider the initial conditions for three targets and additional simulation parameters to be the same for all transmitter-receiver configurations.







\subsection{Target tracking}
Localization and tracking in MIMO radar depend on the target's initial states and the geometry of the system \cite{9364752}. Target tracking scenarios using the L-shape array configuration are shown in Figure~\ref{fig:Lshape}, using Circular and Random array configurations can be found in Figure~\ref{fig:circle} and Figure~\ref{fig: random_pdf.pdf}, respectively. All these three figures suggest that in a distributed MIMO radar system, the circular array configuration is the most efficient in tracking the targets among the three array configurations. This observation agrees well with the results in Figure~\ref{fig:Pc}, where the probability of correct association has been investigated. This is because both performance metrics complement each other. Table \ref{tab:2} demonstrates the echo sequence for each target from the corresponding transmitter receiver for the circular transmitter-receiver configuration.
Each receiver receives multiple echoes, which are unordered in nature. From the estimated target location using the JPDA algorithm, we estimate the order of echo at any receiver.
\begin{table}[t]
\centering
\caption{Echo Sequence from three targets, before and after association.}
\resizebox{\textwidth}{!}{%
\begin{tabular}{|c|c|c|c|c|c|}
\hline
  & $\mbox{Tx}_1$  & $\mbox{Tx}_2$ & $\mbox{Tx}_3$ & $\mbox{Tx}_4$ & After Association \\ \hline
$\mbox{Rx}_1$  & target 3,2,1 & target 3,2,1 & target 2,3,1 & target 1,3,2 & 3,2,1 \\ \hline
$\mbox{Rx}_2$  & target 3,2,1 & target 3,2,1 & target 2,3,1 & target 1,3,2 & 3,2,1  \\ \hline
$\mbox{Rx}_3$  & target 3,2,1 & target 3,2,1 & target 2,1,3 & target 1,2,3 & 1,3,2 \\ \hline
$\mbox{Rx}_4$  & target 3,1,2 & target 3,2,1 & target 1,2,3 & target 1,2,3 & 1,3,2 \\
\hline
\end{tabular}}
\label{tab:2}
\end{table}
Here, we notice that the joint association probability helps successfully associate each measurement with the corresponding targets.

\section{Conclusion}
This is a concluding paper of a three-part series. The first two companion papers Part I \cite{liu2024codesigningpart1} and Part II \cite{liu2024codesigningpart2} investigated, respectively, the signal processing and the synergistic design algorithm for a IBFD MU-MIMO communications system that shares spectrum with a distributed MIMO radar with a single target in its coverage area. In this paper, we handled the co-design challenge for a multiple target scenario. A method of low computational complexity, leveraging the Barzilai-Borwein gradient algorithm, was introduced to derive the design parameters efficiently. Furthermore, we employed a mixed-integer linear programming approach to facilitate distributed target localization. The feasibility and precision of multi-target sensing capabilities within the distributed IBFD ISAC framework were validated through comprehensive numerical experiments. Additionally, our study showcased the practical application of the IBFD MU-MIMO communication system and distributed radar ISAC system for localizing and tracking multiple targets, employing advanced techniques such as the JPDA and extended Kalman filter. Three different transmitter-receiver array (L-shape, circular, and random) configurations were considered. Among those, the circular configuration exhibited the best association and tracking performance because of its $360^\circ$ coverage of the surveillance area. 


\label{sec:conclusion}

\bibliographystyle{elsarticle-num}
\bibliography{SP_SI_Part2_v01}

\end{document}